\documentclass[useAMS]{mn2e}  

\usepackage{graphicx}	
\usepackage{amsmath}	
\usepackage{amssymb}	
\usepackage{subfig}
\usepackage{multicol,caption}

\def\lesssim{\mathrel{\hbox{\rlap{\hbox{\lower4pt\hbox{$\sim$}}}\hbox{$<$}}}}
\def\gtrsim{\mathrel{\hbox{\rlap{\hbox{\lower4pt\hbox{$\sim$}}}\hbox{$>$}}}}
\newcommand{\ltsima}{$\; \buildrel < \over \sim \;$}
\newcommand{\simlt}{\lower.5ex\hbox{\ltsima}}
\newcommand{\gtsima}{$\; \buildrel > \over \sim \;$}
\newcommand{\simgt}{\lower.5ex\hbox{\gtsima}}

\title[Second generation stars in GCs]{Formation of Second Generation Stars in Globular Clusters}

\author[F. Calura et al.]{
F. Calura$^{1}$\thanks{E-mail: francesco.calura@inaf.it}, A. D'Ercole$^{1}$, E. Vesperini$^2$, E. Vanzella$^{1}$, A. Sollima$^{1}$\\
$^1$ INAF - OAS, Osservatorio di Astrofisica e Scienza dello Spazio di Bologna, via Gobetti 93/3, I-40129 Bologna, Italy\\
$^2$ Department of Astronomy, Indiana University, Bloomington, IN 47401, USA\\
}

\date{}

\pubyear{}

\begin{document}
\label{firstpage}
\pagerange{\pageref{firstpage}--\pageref{lastpage}}
\maketitle

\begin{abstract}
 By means of grid-based, 3D hydrodynamical simulations we study the
formation of second generation (SG) stars in a young globular cluster
(GC) of mass $10^7~M_{\odot}$, the possible progenitor of an old GC with a
present mass $\sim~(1-5)~10^6~M_{\odot}$. The cluster accretes external gas as 
its first generation (FG) asymptotic giant branch (AGB) stars release
their ejecta and SG stars form. We consider two models characterised
by different densities of the external gas. In both cases, we find
that a very compact SG subsystem with central density  $>10^5~M_{\odot}/pc^3$ 
forms in the innermost regions of the cluster. The low-density model
forms a population of extreme SG stars with high helium enhancement, 
followed by the formation of another SG group out of a mix of pristine
gas and AGB ejecta and characterised by a modest helium enhancement. 
On the other hand, the high-density model forms in prevalence SG stars 
with  modest helium 
enhancement. Our simulations illustrate the dynamical processes governing
the formation of SG populations in GCs 
and shed light on the structural properties emerging at
the end of this phase. The newly born SG groups have different concentrations,
with more extreme SG stars 
more centrally concentrated than those with less extreme chemical
abundances. The very high density of the SG subsystems implies that SG 
massive stars, if formed, might suffer frequent close encounters,
collisions and gas stripping, thus possibly contributing further gas
to the SG formation.
\end{abstract}

\begin{keywords}
Hydrodynamics; methods: numerical; globular clusters: general; galaxies: star formation. 
\end{keywords}



\section{Introduction}
\label{sec_intro}
In the last decades, strong evidence of the presence of multiple
stellar populations has been found in all the Galactic GCs that have
been investigated by means of photometric or spectroscopic studies. 

Many  spectroscopic studies revealed that GCs host stellar populations
that are not chemically homogeneous; even if GC stars are generally
homogeneous in their Fe content, they are characterised by significant
star-to-star abundance variations in a number of light elements, such
as C, N, O, Na, Mg, Al (see e.g. Carretta et al. 2009a,b, Gratton et
al. 2004, 2012 and references therein).  More recently, high-precision
photometric observations have provided further evidence in support of
the presence of multiple stellar populations in GCs (see e.g. Lee et
al. 1999, Ferraro et al. 2004, Bedin et al. 2004; Piotto et al. 2005,
2007, 2015; Marino et al. 2008, Milone et al. 2017; for a review, see Gratton et al. 2012; Bastian \& Lardo 2018). 

Despite the wealth of observational data collected so far, 
our current understanding of the origin of multi-populations in GCs is
still incomplete. 
Several models have been proposed so far, none of which is free from drawbacks and assumptions invoking complex
physical conditions sometimes difficult to test. 
One of the most thouroughly studied model is the asymptotic giant branch (AGB) scenario, 
in which a second generation (SG) of stars forms from the ejecta of first generation (FG)
AGB stars (e.g. D'Ercole et al. 2008, 2010, 2012, D'Antona et al. 2016).
Other scenarios proposed so far include models in which SG stars form
from gas processed in fast-rotating massive stars (Decressin et al. 2007), massive interacting binaries (de Mink et al. 2009 ), black hole accretion discs
(Breen 2018), massive (Elmegreen 2017), or even supermassive ($M > 10^3~M_{\odot}$) stars
(Denissenkov \& Hartwick 2014; Gieles et al. 2018). 
All scenarios need to face a variety of complex observational constraints concerning, for example, the chemical properties of multiple populations,
the number of populations, the relative number of FG and SG stars, as well as their dynamical properties.
One of the constraints often discussed in the literature concerns the initial mass of the FG cluster needed to produce the observed number of SG stars.
All the possible sources of processed gas for the SG formation proposed in the literature do not release large quantities of gas and, 
to account for the fraction of SG stars observed in GCs today, it is therefore necessary to assume that the mass of the GC precursors had to be initially larger
than present-day clusters. 

The hydrodynamical study presented here is conducted within the AGB
scenario and extends the early exploration by D'Ercole et al. (2008). 
By means of 1-D hydro simulations, D'Ercole et al. (2008) showed that the
AGB ejecta collects in a cooling flow leading to the formation of a
more centrally concentrated SG subsystem in the central regions of the
primordial cluster. Several observational studies have found clusters
in which some memory of this initial spatial difference is still
retained and in which the SG is more spatially concentrated than the
FG (see e.g. Sollima et al. 2007, Bellini et al. 2009, Lardo et
al. 2011, Milone et al. 2012, Simioni et al. 2016).  The possible
origin of the chemical abundances of elements such as Na, O, Al, Mg,
Li in multiple population clusters have been extensively studied in
the AGB scenario (see e.g. D'Ercole et al. 2010, 2012, D'Antona et
al. 2016, D'Antona et al. 2019) along with the possible origin of
discrete groups in the SG populations. We refer to those papers for  a
detailed discussion of the uncertainties and the various constraints
and open questions concerning the AGB stellar models nucleosynthesis.
A problematic aspect pointed out in these studies concerns the fact
that in order to account for the chemical composition of GC stars, in
particular the pattern followed by light elements, the AGB scenario
must invoke dilution of AGB ejecta with pristine gas occurring during
SG formation. A similar requirement of dilution with pristine gas is
shared by other models (see e.g. Decressin et al. 2007, Denissenkov et
al. 2015, Gieles et al. 2018). The requirement of pristine gas in the
mix out of which SG stars form opens a new fundamental question
concerning the hydrodynamics of this gas during the SG formation and
provides the main motivation for the study presented in this paper. An
additional problem which is briefly discussed in this paper but which 
will require further exploration in future studies concerns the
possible role of feedback of SG massive stars and other energy
sources.

Despite the general consensus regarding the need for dilution, the
source of the diluting pristine gas is still not clear, and all the
mechanisms proposed so far (Bekki \& Mackey 2009, Pflamm-Altenburg \&
Kroupa 2009, D'Ercole et al. 2010; 2011, Gratton \& Carretta 2010,
Conroy \& Spergel 2011, D'Ercole et al. 2016) require further
explorations of the gas dynamics and the implications for the
resulting abundance patterns. 

Because of the complexity and the variety of the physical processes involved in the accretion process, 
an analytic estimate of the mass of  external gas accreted by a cluster is  problematic. 
While for a point mass accretor the 
mass accretion rate can be estimated by means of the Bondi-Hoyle-Lyttleton 
formula (Bondi 1952), an analogous simple analytical formula for 
the accretion onto potentials for extended mass distributions is not available, and one must therefore 
resort to numerical simulations.  
A significant step forward in this field is represented by the work of Naiman et al. (2011; see also Kaaz et al. 2019), who  performed
a set of numerical simulations appropriate for different astrophysical contexts. 
Naiman et al. (2011) modelled the accretion onto various 
core potentials (e. g. of the Plummer- and Hernquist-type) moving through a background medium, as example studies of 
various astrophysical environments, including 
gas accretion onto star clusters. 
The results of their simulations shed light on a number of key aspects of the accretion process although
they still invoke a few important simplifications,
in that mass return from cluster stars, the self-gravity of the gas, and star formation are not included. 
The simulations of Naiman et al. (2011) show that, in general, the accumulation
of external gas occurs during an initial period lasting between $\sim$10 and 100 sound
crossing times of the cluster. Their results indicate that eventually, a steady configuration is
established, and the gas accumulated into the cluster ceases to grow
appreciably. 
According to their results, the gas
accumulated into the core may reach very high densities and, since the
local free-fall time is very short, new stars could, in principle form rather quickly. 
Although this is a very promising result as it suggests that the
cluster can indeed re-accrete the correct amount of needed pristine gas (see e.g. D'Ercole et al. 2010, 2011, 2012, 2016), 
in order to have a more realistic physical description of a proto-GC, additional simulations are necessary.
In particular it is important to consider simulations combining the mass return from  AGB stars,
the accretion of pristine gas, the gas self-gravity 
as well as the formation of SG stars, four 
ingredients which have never been simultaneously taken into account in previous numerical studies of GC formation.

In this paper, we present for the first time three-dimensional hydrodynamic simulations of
a star forming GC which includes all such ingredients. 
In particular, we investigate various aspects related to the formation and early evolution of a GC in the framework of the AGB scenario
presented in D'Ercole et al. (2016). 
In our simulations, the GC moves through a distribution of external gas and is allowed to accrete mass from it. 
 
Although the setup is still idealised, our numerical experiments are
designed to provide a well controlled, theoretical framework which
includes several key ingredients for the formation of SG stars in a
young GC. We point out that the main goal of the paper is to address a
very specific question concerning the possible reaccretion of the
pristine gas, explore the dynamics of multiple population formation
from this pristine gas and AGB ejecta, and the emerging structural
properties of the multiple population subsystems. Many different
aspects concerning the chemical and dynamical evolution  of multiple
populations which are beyond the scope of this investigation remain to
be further studied. Moreover it will be important to extend the study
of the hydrodynamics of the SG formation in the context of the various
other scenarios proposed in the literature and further explore their
signatures in the present-day chemical and structural properties of
GCs (see e.g. Howard et al. 2019, Bekki 2019).

This paper is organised as follows.
In Section 2 we present our model, in particular the  setup of our hydro-simulation and
describe our most crucial assumptions. 
In Section 3, we present our results and in Sect. 4 we discuss their most important implications
 and future directions. 
Finally, in Sect. 5 we draw our conclusions.

\section{Simulation Setup}
In this work, the initial conditions coincide with
the final stage of the simulations presented in Calura et al. (2015),
although some of the model parameters are different. 
In our initial setup, the stellar first generation is $\sim~t_{AGB}=39$ Myr old. 
At this time, all FG type II Supernovae (SNe) have exploded,
no primordial gas or supernova (SN) ejecta are left in the cluster, 
and the most massive FG AGBs are starting to return mass and energy into the gas-free system. 

As discussed in Sect.~\ref{sec_intro},  
previous studies have shown that the SG cannot form out of 
AGB ejecta only, as the observed abundance pattern followed by light elements 
would not be reproduced 
without a certain amount of dilution from pristine gas
(i.e. not polluted either by FG SNe or by AGB stars;  
see, e. g., D'Ercole et al. 2011; Bastian et al. 2015). 

Our numerical simulations are designed to describe the motion of 
a cluster with respect to a background gas distribution. 
Starting from a given time $t_w>t_{AGB}$, 
pristine gas starts to flow inside our computational box from one of the boundaries. 
In principle, this setup is designed to recreate the physical
conditions described in D'Ercole et al. (2016), in 
which the GC is assumed to lie within the disc of a galaxy at high
redhift (Kravtsov \& Gnedin 2005; Kruijssen 2015). 
However, the gas distribution may also represent other real  
cases such as, e. g., a reservoir of gas leftover from a merger, 
and the motion could also describe a newly born cluster orbiting around the centre of a dwarf galaxy or
migrating towards it (Goodman \& Bekki 2018).

Following D'Ercole et al. (2016), assuming that the FG type II SNe carved a large cavity
into the external gas distribution and that the  FG cluster is moving along it,
the system would experience an asymmetric re-accretion of gas from the side towards which
the cluster is moving.

In order to create such a condition, 
for the purposes of simplicity, we assume a reference frame
in which the cluster is at rest, and we assume that it is
accreting mass from one of the boundaries, as shown in Fig. 1.

For the computation of $t_w$, we follow D'Ercole et al. (2016).
For a cluster in motion with respect to the ISM, the radius at which the ram pressure due to the supernova-driven wind balances
the pressure of the external medium can be computed as: 
\begin{equation}
  R_{eq} [100~pc] = 41.43~\left(\frac{L_{41}}{n_0~V_{w,8}~(\sigma^2_{0,6}+v_{pg,6}^2)}\right)^{\frac{1}{2}}.
\end{equation}
The quantity $L_{41}$ is the mechanical luminosity of the wind in units of $10^{41}~erg~s^{-1}$,
whereas $V_{w,8}\sim 2$ is the velocity of the wind in units of $10^8~cm~s^{-1}$. 
For a cluster of mass $M_{FG}=10^7~M_{\odot}$ and for a standard stellar initial mass function (IMF; i.e. Kroupa 2001; Chabrier 2003), 
we have $\sim10^5$ SNe and $L_{41} \sim 1$. The two terms 
$\sigma_{0,6}\sim 1$ and $v_{pg,6}=2$ are the isothermal sound speed and the velocity of the pristine gas relative to the GC
(both in units of $10^6~cm~s^{-1}$), respectively, whereas $n_0$ is the gas density in $cm^{-3}$. 
The time at which the infall starts can then be computed as
\begin{equation}
  t_w = t_{SN} + \frac{R_{eq}}{\sigma_0 + v_{pg,6}},
\label{eq_tw}
\end{equation}
where $t_{SN} \sim 30$ Myr is the time after which all FG type II SNe have exploded. 
Assuming $V_{w,8}\sim2$ and $n_0\sim \rho_{pg}/m_p$ (where $m_p$ is the proton mass), we obtain $t_w \sim 60~Myr$ and  $t_w \sim 40~Myr$
for $\rho_{pg}=10^{-24}~g~cm^{-3}$ and $\rho_{pg}=10^{-23}~g~cm^{-3}$, respectively.

\begin{figure*}
   \includegraphics[width=0.7\textwidth]{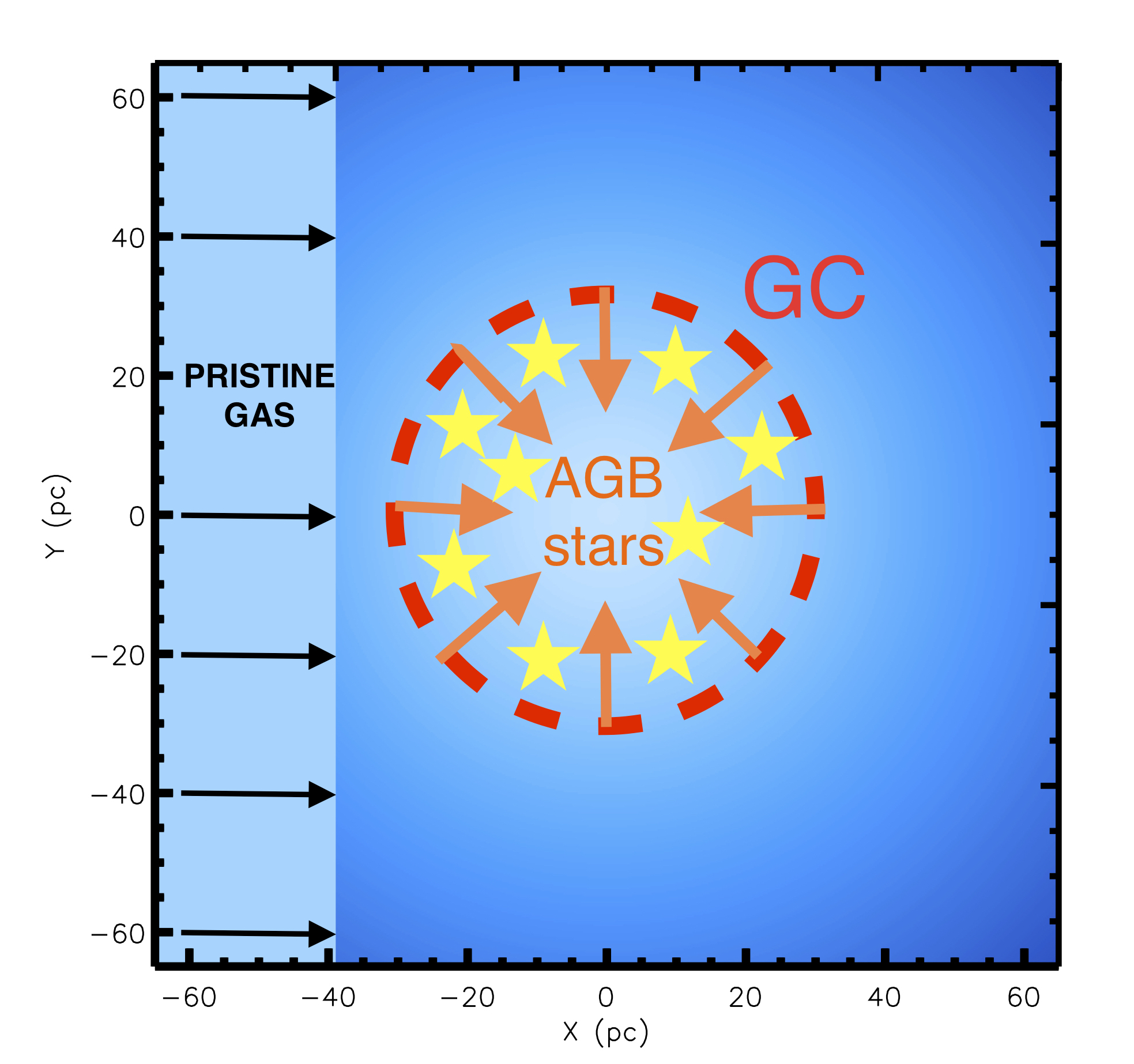}
   \caption{A schematic description of the setup of our simulation. At the beginning of our simulation, corresponding to $t\sim~40$ Myr after the birth of the cluster, 
     first generation AGB stars start releasing their ejecta and second generation stars can form out of this gas. 
     Starting from a given time $t_w$, a uniform pristine gas distribution
with the same chemical comoposition as FG stars enters the computational box from one of the boundaries.} 
        \label{fig1}
\end{figure*}
The cluster is assumed to move with respect to the external uniform gas distribution, with a velocity of the order of
the isothermal sound speed of a medium characterised by a temperature of $\sim 10^4$ K (see Sect.~\ref{sec_thermal}). 
We point out that the computation of $t_w$ requires assumptions regarding several uncertain parameters,
including the cluster velocity with respect to the ISM, the SN mechanical luminosity and the efficiency with which SN
energy is deposited in the ISM, the motion of the cluster with respect to the external gas distribution,
the clumpiness of the medium or the possible presence of density gradients. 
For these reasons, we regard the value computed above for $t_w$  as representative order of magnitude estimate. 
A summary of the main parameters describing the two models considered here is presented in Table 1.

\begin{table*}
\caption{Parameters describing the simulation setup}             
\label{param}      
\begin{tabular}{ l l l }     
\hline
Parameter    & Description & Adopted values \\
\hline
$\rho_{pg}$      &     density of pristine gas  &  $10^{-23}; 10^{-24}~g/cm^{3}$\\
$v_{pg,6}$         &   pristine gas velocity   &  $20~km/s$ \\
$T_{pg}$         &    temperature of the pristine gas & $10^4~K$ \\
$T_{floor}$  &        Minimum temperature of the simulations & $10^3~K$ \\
$t_*$         & Star formation timescale            & 0.1 Gyr\\
$M_{FG}$    & Mass of first generation stars & $10^7~M_{\odot}$\\ 
$r_{Plum}$   & Plummer radius of FG stellar distribution & $23~pc$\\ 
$t_{end}$    & Duration of SG star formation  & $65~Myr$\\
\hline 
\end{tabular}                                                              
\end{table*}
\label{tab_param}
We performed several low-resolution tests in which we varied
the basic parameters describing the flow of matter onto the cluster.
In particular, attempts carried on with a density of the infalling medium by more than 
one order of magnitude larger than the one assumed here (i.e. $>$10 cm$^{-3}$ or more)
provided negative results for the capability of the system to retain the gas, in that SG stars  
were originating in a loose, unbound structure rapidly dissolved by the flow,
hence could not aggregate into a bound core.
A similar outcome was obtained by assuming a flow velocity much larger than the value adopted here, which roughly corresponds to 
the double of the isothermal sound speed of a gas at $\sim~10^4$ K. 
In fact, as also shown by a few previous studies, in such conditions an efficient accretion may
work only for a limited range of values of the parameters of interest.
If the energy of the incoming gas is much larger than the
GC potential well, a collective accretion is prevented. Moreover,
if the cluster velocity is too large, the ram pressure exerted on the
gas at the bottom of the potential well becomes stronger than the
gravitational restoring force, and the sweeping process clears out
the cluster of its gas instead of promoting accretion (Mori \& Burkert 2000; Lin \& Murray 2007; Conroy \& Spergel 2011). 

The physical conditions of the infalling material in the simulations on which this paper is focused, which are reported in Table~\ref{tab_param}, are not far from
those observed in real systems, such as the gas density in a dwarf galaxy (Marcolini et al. 2003 and references therein)
or in a high redshift disc (D'Ercole et al. 2016; Wardlow et al. 2017).

\subsection{Main physical ingredients}
In this work, we use a customized version of the RAMSES hydrocode
(Teyssier 2002) to study star formation occurring within a young GC by means of 3D hydrodynamic simulations. 
Besides mass and energy return from 
AGB stars, our model also includes radiative cooling 
and star formation. The gravitational effects of FG and SG stars and 
the self gravity of the gas are taken into account, along with 
the chemical composition of the stellar ejecta and of newly formed stars.
In this section we describe the main ingredients of our hydrodynamic simulations. 
The total, cubic computational volume is  $L_{box}^3=160^3 pc^3$. 
Our simulations are run using a uniform grid, with maximum resolution equal to $\sim 0.3~pc$. 
The gas evolution is computed using a second-order Godunov scheme 
for the Euler equations. Collisionless star particles are allowed to form, and their 
trajectories are computed by means of a Particle-Mesh solver.

The gravitational effects of FG stars are taken into account by means of a static
potential, computed from their density profile (eq. 5). 
The self-gravity of the entire system composed by SG stellar particles and
gas is fully taken into account, and these two components do exert gravitational
force on each other and are subject to the effects of FG stars.

\subsection{Mass, energetic and chemical feedback from AGB stars}
Mass return from AGB stars is modelled by adding a source term
to the mass conservation equation, representing the rate of mass
return per unit volume and expressed as:
\begin{equation}
\dot \rho_{*,AGB} = \alpha \rho_{*}, 
\end{equation}
where $\rho_{*}$ is the stellar density of FG stars. 
The quantity $\alpha$ is the specific mass return rate and is computed following
the formalism of Ciotti et al. (1991) (see their equation 3) adopting a   
universal IMF (Kroupa 2001).
If $t_{yr}$ represents the time expressed in yr, 
the quantity $\alpha$ as a function of time (expressed in $yr^{-1}$) 
is represented by the analytical formula 
\begin{equation} 
  \alpha(t_{yr}) = 0.065 \cdot t_{yr}^{-1.01}. 
\label{alpha}
\end{equation}
The formula is valid for times between 30 Myr and 100 Myr. 

FG stars follow a Plummer (1911) mass density profile: 
\begin{equation} 
\rho_*(r) = \frac{3M_{tot}}{4\pi\, a^3} \left(1+\frac{r^2}{a^2}\right)^{-\frac{5}{2}}, 
\label{plum}
\end{equation}
with $M_{tot}=10^7 M_{\odot}$ and $a=23$ pc (corresponding to a half-mass radius equal to 30 pc).
The mass chosen is meant to represent that of the possible progenitor of the most massive Galactic GCs 
(after mass loss occurring during the cluster's early and long-term dynamical evolution is taken into account; see e.g. D'Ercole et al. 2008).
We emphasize that the characteristic radius and the stellar mass for FG stars chosen here are very similar to the values observed in proto-GC candidates at
z$\sim$3 and z$\sim$6, recently detected in gravitationally lensed fields (Vanzella et al. 2017a,b).

We emphasize that the stellar mass of the FG enclosed within the computational box adopted for the simulations includes $\sim 90\%$ of $M_{tot}$
and, therefore, for the purposes of calculations concerning the relative SG to FG mass in the rest of the paper,
the mass of the FG system to be considered is equal to $0.9~M_{tot}$ $(9 \times 10^6~M_{\odot})$. 

The energetic input from FG AGB stars is taken into account by means of the source term:
\begin{equation} 
S = 0.5~\alpha~\rho_*~(3\sigma^2 + v^2 + v_{wind}^2), 
\end{equation} 
where $v$ is the fluid velociy, $\sigma$ is the velocity dispersion,
computed for a Plummer sphere following Dejonghe (1987), 
and $v_{wind}\sim 2~10^6~cm~s^{-1}$ is the wind velocity for AGB stars (D'Ercole et al. 2008). 
The source term $S$ represents the heating of the gas due to the feedback from FG AGB stars and 
is added at each timestep to the fluid energy.

\subsubsection{Thermal modelling}
\label{sec_thermal}
Besides heating from AGB stars, our simulations also include radiative gas cooling. 
The cooling function implemented in RAMSES takes into account 
H, He and metal cooling (Sutherland \& Dopita 1993, see Few et al. 2014).
The contribution from metals at temperatures above $10^4$ K
is accounted for through a fit of the ratios between the 
cooling rates calculated at solar metallicity and those at zero
metallicity using the photoionisation code CLOUDY (Ferland et al. 1998).

At lower temperatures, metal line-structure
cooling rates are from Rosen \& Bregman (1995).

The temperature chosen for the external medium, i.e. the pristine gas which continously 
enters the computational box from one of the boundaries, is $10^4$~K.
Such a temperature value is quite standard for the
interstellar medium of a 'normal', star-forming 
galaxy which is kept photoionised by a stable UV stellar radiation field (e. g., Haffner et al. 2009).

The temperature threshold for radiative cooling adopted in this work is $10^3$ K,
i. e. anywhere in the volume the gas is not allowed to cool down to a temperature below this value.
This choice is motivated by the impossibility of resolving the
Jeans length, which decreases with the minimum temperature, set by the
temperature floor. 
With our assumption ($T_{floor}=10^3$ K) and with a density of $\sim 10^{-16} g~cm^{-3}$ (the maximum achievable in
clusters), the Jeans length is of the order of 0.01 pc (and it would be even smaller for a lower value of $T_{floor}$),
well below our maximum resolution. 
In general, in our simulations numerical convergence is satisfactory
(see Fig.~\ref{fig_mass}) and we do not find any 'density
runaway', which sometimes may occur when the Jeans length is not well-resolved (Truelove et al. 1997). 
Moreover, at very large densities, in principle a lower value for $T_{floor}$ 
should not lead to strong variations in our results, as the cooling time is
generally much lower than the Courant time, and the amount of
gas that rapidly cools to form stars is not expected to change much. 

Our simulations are aimed at
roughly modeling a two-phase medium (i.e. a ’neutral’ one, where star formation can occur,
and a ’ionised’ ambient which surrounds the star-forming centre of the cluster). 
Our equation of state is $P \propto \rho^\gamma$, where $P$ and $\rho$ are the pressure and density of the gas, respectively, and $\gamma=5/3$. 

\subsubsection{Star formation model}
\label{sec_sf}
The star formation (SF) model used in this work relies upon the native
RAMSES SF implementation as described in detail in Rasera \& Teyssier (2006).

In each cell eligible for star formation, the gas  
can be converted into star particles and new stars with density ${\rho}_{*,SG}$ can
form according to:
\begin{equation}
  \dot{\rho}_{*,SG} = \frac{\rho}{t_*}, 
\end{equation}
i.e. according to the Schmidt (1959) law.
In each cell, the star formation timescale $t_*$ is proportional to the local free-fall time, and computed as
\begin{equation}
  t_{*}=  t_0 \left( \frac{\rho}{\rho_0} \right)^{-1/2}.
\end{equation}
Here we have combined the unknown quantities $t_0$ and $\rho_0$ in order to have 
$t_* \sim 0.1$ Gyr for a density $\rho$ of the order of 1 atom per cubic centimeter. 
However, our results do not depend significantly on the choices of the free parameters $t_0$ and $\rho_0$. 
As shown by D'Ercole et al. (2008), this is true when the replenishment of the gas occurs
on timescales which are much shorter than the timescale of the simulation. 

For numerical reasons, the code allows that no more that 90\% of the gas in the cell can be used for star particle formation. 
A single star particle is formed in each cell cooled down to a temperature $T<2 \cdot 10^4$ K (i. e.
where the gas is assumed to be neutral) and where the net flow is converging, i. e. in which
\begin{equation}
(\nabla \cdot \mbox{\boldmath{$v$}}) < 0,
\end{equation}
where $\mbox{\boldmath{$v$}}$ is the fluid velocity in the cell.

The mass of each collisionless stellar particle is $M_p= N~m_0$, where $m_0=0.1~M_{\odot}$ is the minimum mass and $N$ is
determined by sampling from a Poisson distribution, characterised by a probability $P$ given by
\begin{equation}
P(N)={\lambda_P \over N !} \exp({-\lambda_P}) \, ,
\label{poisson_law}
\end{equation} 
where the mean value is calculated as: 
\begin{equation}
\lambda_P= \left ( {\rho \Delta x^3 \over m_0}\right ) {\Delta t \over t_*}. 
\label{poisson_param}
\end{equation}
In this equation $\Delta t$ represents the timestep, whereas $\Delta x$ is the cell size.

These star particles are placed at the centre of their parent cell, with a velocity equal to
the local fluid velocity. 
The corresponding mass, momentum, energy and density in the form of the chemical element $k$
are conservatively removed from the ones of the parent cell.

Star formation is halted after 65 Myr,  
corresponding to $t_{AGB}+65~Myr=104~Myr$ after the
formation of the FG.  This choice for the duration of the SG star
formation event is determined by the uncertain timescale of the explosions of
the first FG Type Ia SNe. Our choice is consistent
with the timescales suggested by observational studies (see
e.g. Mannucci et al. 2006; Maoz et al. 2014) but is meant to provide
just a general indication of the timescale of the SG star formation event.
The implications of 
earlier explosions of SNIa are illustrated by the panels in Figures \ref{fig_chem1} and \ref{fig_chem1_densew},
showing the results at earlier times, whereas later SNIa explosions would
further extend the SG star formation episode possibly increasing the
fraction of SG stars with very moderate chemical anomalies. A more
extended SG formation would also imply a contribution 
of low-masses AGB ($m<5M_{\odot}$) possibly
producing SG populations with the s-process elements anomalies
observed in some clusters (see e.g. D'Antona et al. 2016 for a further
discussion).  Energy release from SG massive stars, if they form,
could provide another mechanism halting SG star formation (see
Sect.~\ref{sec_ms}). We note that in the models
presented in this paper star formation is continuous, but in reality
it could be more complex and characterised by separate star formation
bursts (see e.g. D'Antona et al. 2016 for a discussion of the possible
role of delayed SNII events). Modeling SG star formation including
separate bursts, the detailed implications of different durations for
the star formation event, and the processes behind them is beyond the
scope of the present work and will be addressed in a future study. 

Finally, as far as the chemical properties of SG stars formed are
concerned, we follow only the evolution of the helium abundance. 
The helium abundances used in this work are
those calculated in Ventura \& D'Antona (2011; for the helium yield of
the 8 $M_{\odot}$ progenitor we adopt a value approximately equal to
the average of the yields of the model of Ventura \& D'Antona and 
by Siess (2010); see Ventura \& D'Antona 2011), calculated for a metallicity $Z=10^{-3}$. 
The focus
of the present study is only on the dynamics of SG formation and we will 
not go into any detailed characterization of the abundance patterns of
other elements such as Na, O, C, N, Al, Mg. The study of the origin of
the observed abundance pattern and the nucleosynthesis of the
possible polluters has been the subject of many studies and is still the
subject of intense and debated observational and theoretical
investigations (see e.g. D'Ercole et al. 2010, 2012, D'Antona et
al. 2016, Denissenkov \& Hartwick 2014, Denissenkov et
al. 2015). The high complexity of the questions concerning the
chemical properties of SG stars clearly show the need for further
specific studies of this aspect of the multiple population
phenomenon. In future studies we will expand our investigation to
combine the dynamical study with the chemical aspects and address more
specifically the possible origin of the observed abundance pattern for more chemical elements.  

Finally, one note is in order regarding the chemical composition of
the diluting material which mixes with the AGB ejecta when SF
occurs. The composition of such gas is not expected to be
significantly different from that of FG stars. If the globular cluster
is orbiting in a dwarf galaxy of metallicity $Z \sim 0.001$, the heavy
element content of the gas encountered by the system is not likely to
have changed much on a timescale  of 40 Myr. In fact, chemical
evolution models for galaxies of different morphological types
indicate that significant metallicity variations occur on rapid
timescales only in extremely metal-poor environments (e. g., Calura et
al. 2009). Moreover, radial metallicity gradients in dwarf galaxies 
are typically of the order of $\sim$ 0.1 dex/kpc (see e.g. El-Badry et
al. 2016; Wang et al. 2019). This implies that, even in the extreme case in which the
system was moving on a completely radial orbit with a speed of 20 
km/s, the variation in metallicity would be of 0.08 dex,
i. e. within the metallicity spread (as traced by [Fe/H]) of FG and SG stars in most GCs (e. g., Renzini et al. 2015).
Clearly, a less extreme orbit spanning a smaller range of radial distances would lead
to a smaller metallicity variation.

\section{Results}
In this paper we present the results of models including the
infall from a homogeneous, isothermal gas distribution;
specifically, we focus our attention on a model in which
the infalling gas has a density $\rho=10^{-24}~g/cm^3$,
and another with an infalling gas ten times denser. 

These two models provide an initial framework 
illustrating several key aspects of the dynamics of the SG formation and the link between
the dynamics and the resulting SG chemical properties. 
In both cases, the general behaviour of the gas and of the newly born stellar population 
is first analysed by means 
of two-dimensional density and temperatures maps for the gas distribution. 
For the purposes of clarity, for all the results presented in the remainder of the paper the times will
be expressed assuming $t_{AGB}=39~Myr$ 
as the initial value (e. g., $t= 3~Myr$ will correspond to $t_{AGB} + 3~Myr=42~Myr$ after the formation of FG stars). 
A more extended survey of models exploring different 
properties of the cluster and of the external environment will be matter for future studies.

\begin{figure*}
  \begin{center}
\includegraphics[width = 3.45in]{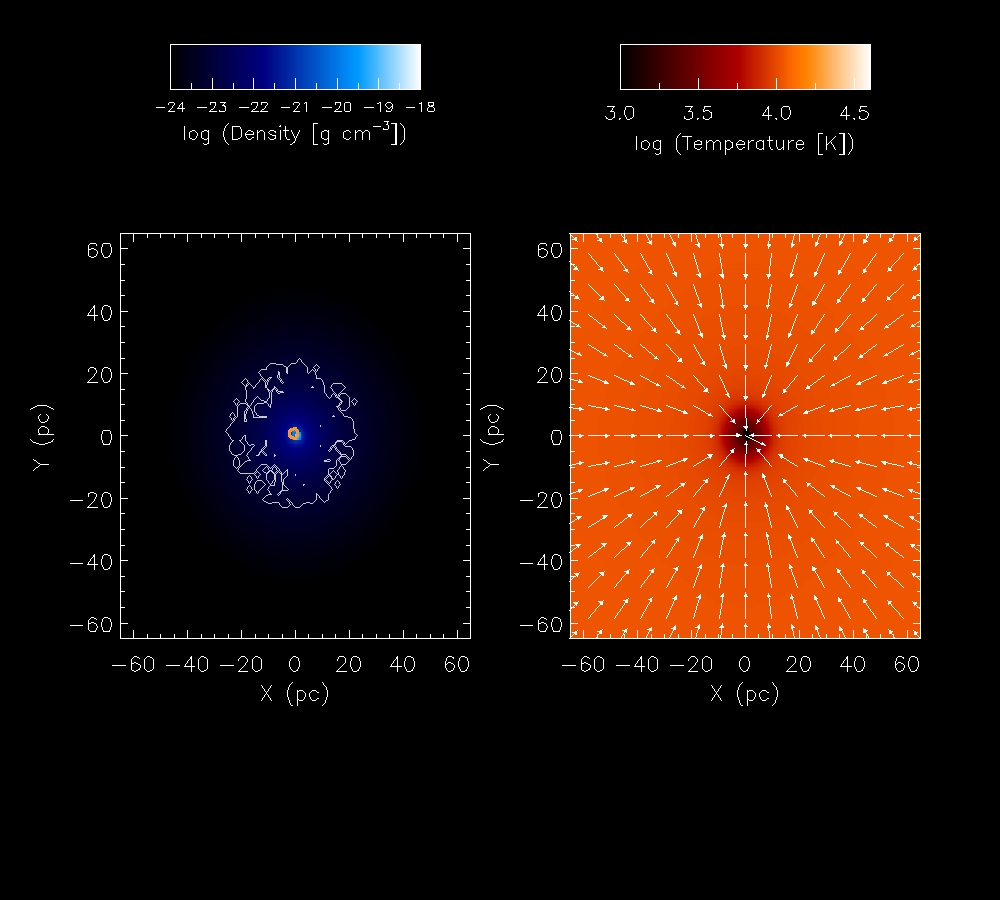}
\includegraphics[width = 3.45in]{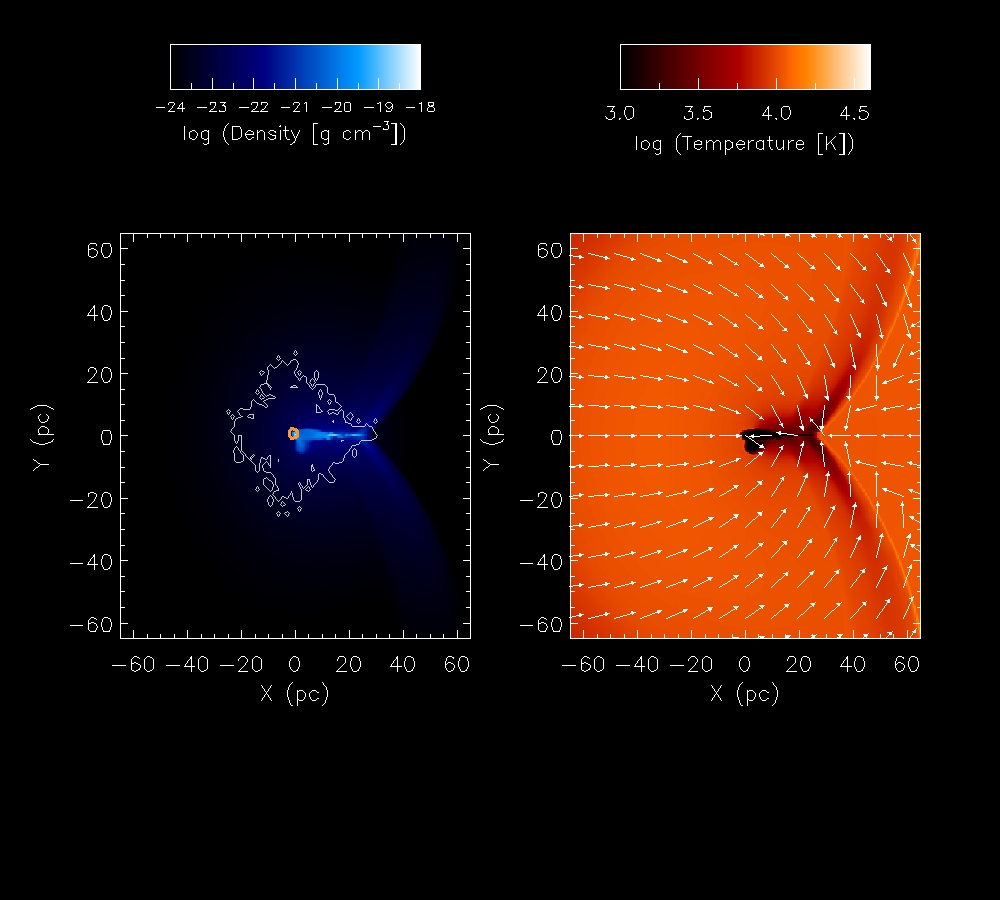}
\includegraphics[width = 3.45in]{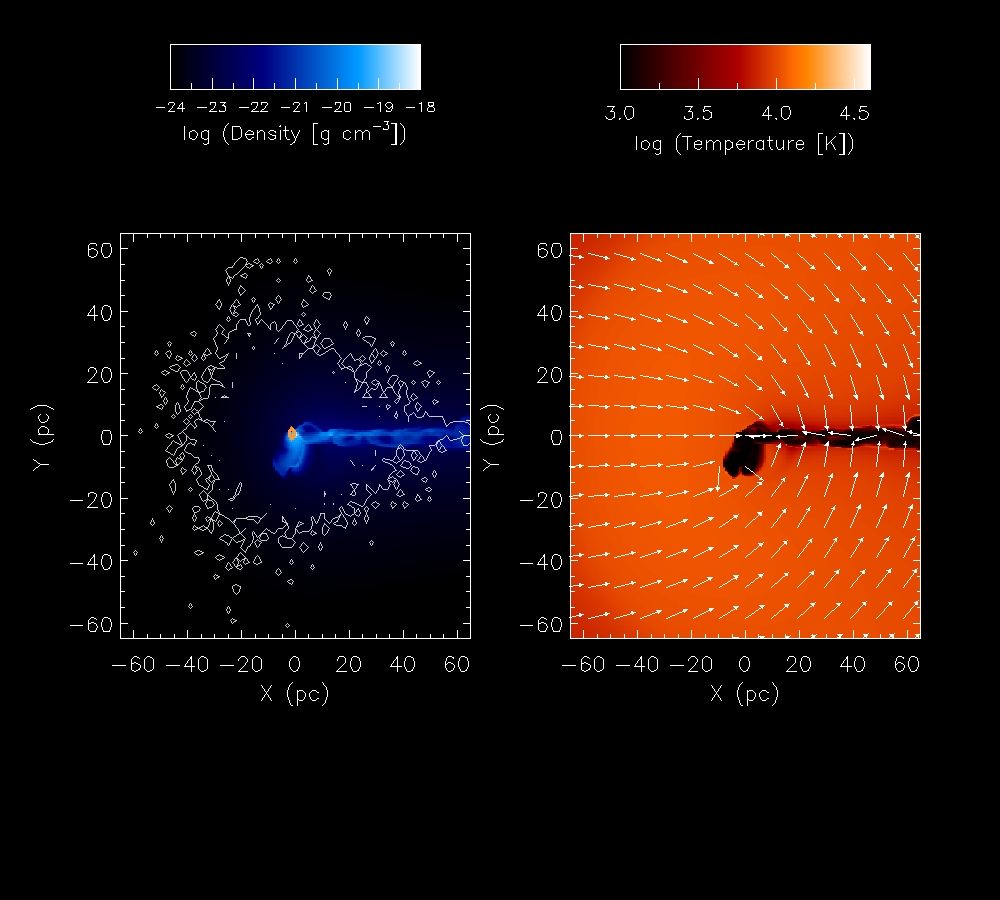}
\includegraphics[width = 3.45in]{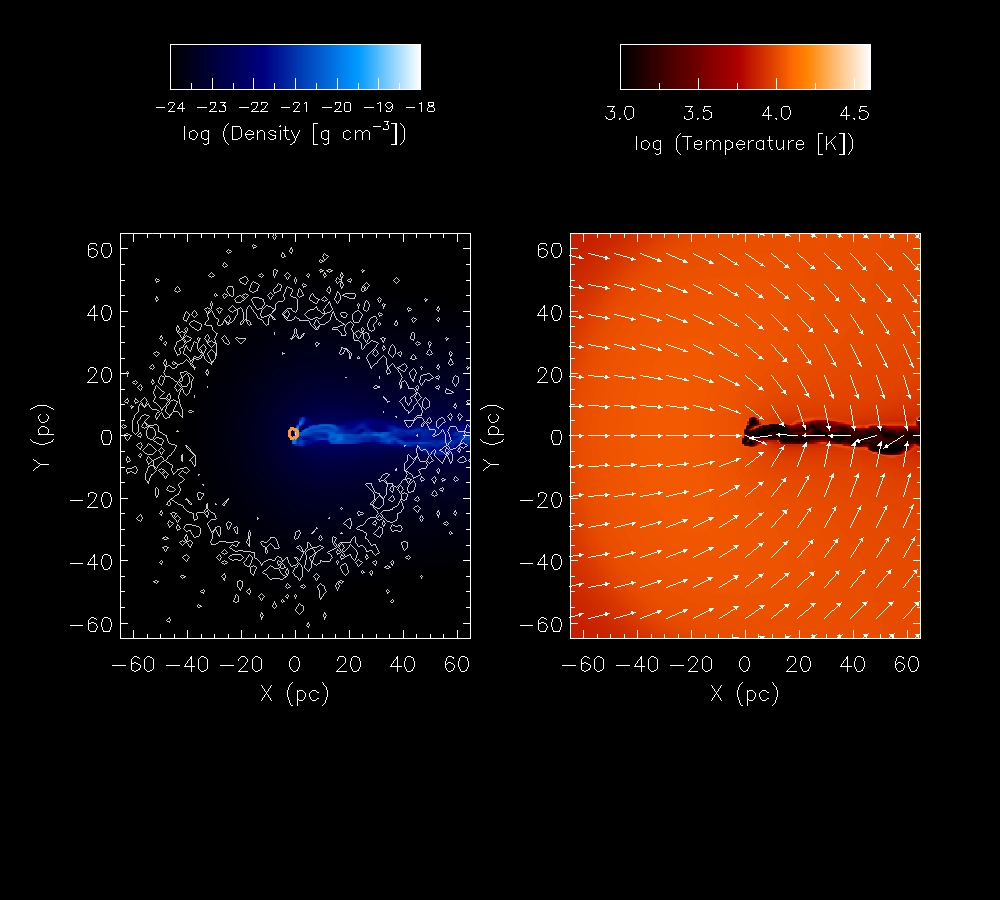}
\end{center}
  \caption{Two-dimensional gas density (left panels of each plot, in black-blue-white colour scale) and temperature (right panels, in black-orange-white colour scale) 
    maps in the {\it x-y} plane in our low-density simulation at various timesteps 
  (from top-left, clockwise: 10 Myr, 26 Myr, 39 Myr, 63 Myr). The white arrows in the temperature maps represent the velocity field.
The white and orange contours describe regions where the SG stellar density is $\ge~10^{-6}$ and $\ge~0.5$ times the maximum value, respectively. The orange contours 
enclose large fractions ($>65 \%$) of the SG stellar mass.}
\label{fig_maps}
\end{figure*}

\begin{figure*}
   \includegraphics[width=0.7\textwidth]{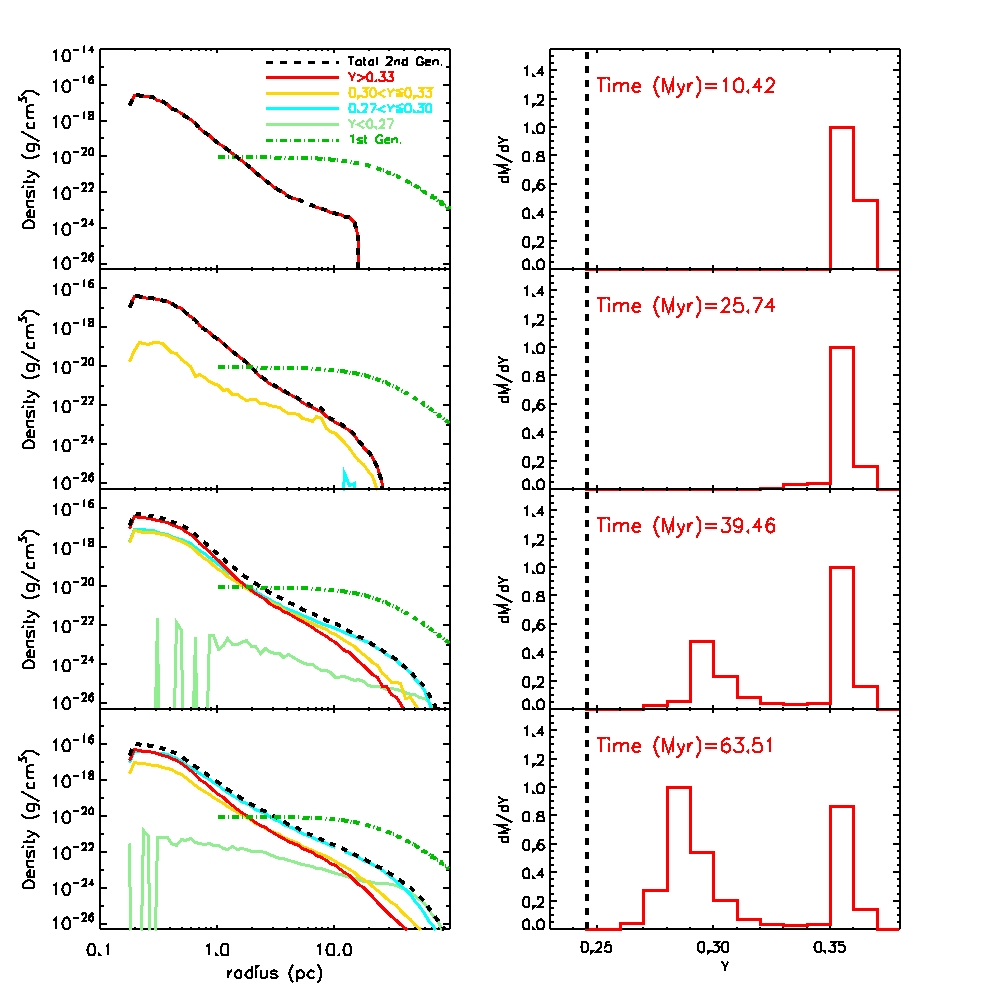}
   \caption{ Left column: total density profile for SG stars at various timesteps and profiles for stars of different He abundance
       for our low-density model ($\rho=10^{-24}~g/cm^3$). 
Also the density profile of FG stars is reported (see the legend in the top-right panel). Right column:
distribution of the He mass fraction Y in SG stars at various timesteps (indicated in each panel). 
The distributions have been computed by adding up the total stellar mass 
in each bin of Y, and then normalizing each distribution to its maximum value. 
The dashed lines represent the values of FG stars.}
        \label{fig_chem1}
\end{figure*}

\subsection{Low-density model}
\subsubsection{Dynamical evolution of the gas}
\label{sec_hydro}

In Fig.~\ref{fig_maps}, we show two-dimensional density and temperature maps
computed at four different evolutionary times of the low-density model run. 
The maps have been calculated by selecting all the grids crossed by a plane
centered at the middle of the computational volume and perpendicular to the {\it z-}axis.  
In this case, the evolutionary times are $t=~10~Myr$, $t=26~Myr$, $t=39~Myr$ and $t=63~Myr$.

At $10~Myr$, the system is still isolated, in that the infall of pristine gas has still to start.

At this time, the ejecta of the AGB are giving rise to a cooling flow, directed towards
the cluster core, where SG stars have begun to form. 
The cooling flow is visible from the velocity field of the gas in the top-left temperature map of Fig.~\ref{fig_maps}, traced by the thin white arrows, all directed towards the centre of the cluster.

At the centre of the box, a cold condensation of gas is already visible, characterised 
by a temperature very close to $T_{floor}$ ($10^3$ K). 
Such a condensation has purely formed out of the matter ejected by AGB stars
and is characterised by a fairly regular, spherical shape. 
In the density map, the thin, white contour encloses the entire SG stellar mass in the  {\it x-y} plane, computed 
considering a reference frame with an origin at the centre of the computational volume. 
As visible from the top-left density map, 
a  small amount of stars has already formed at the very centre of such a condensation. 
The stellar component is extremely compact, as shown by the thick orange contour, 
which at this time encloses more than $90\%$ of the SG stellar mass in the plane. 

At this time, the mass which has been turned into stars is $\sim~10^5~M_{\odot}$, and
it consists of all the most extreme, He-rich stars representing the first SG stars ever formed.

At a later time ($t=26~Myr$), the infall of pristine gas has already started, and 
the incoming shock front has just crossed the centre of the cluster.
Upstream of the cluster, the velocity field reflects the motion of the incoming gas, in that the arrows appear aligned
along the symmetry axis, and otherwise are tilted by the gravitational acceleration caused by the cluster itself,
with an inclination gradually increasing along the {\it y}-axis and as the distance from the centre gets smaller and smaller. 
The symmetric, thin density enhancement caused by the shock is visible from the two 'wings' 
located at the right side of the cluster centre. This is caused by 
the distorted front of the shock, which appears warped into a sharp cusp pointing towards the centre. 
The interaction between the cold AGB ejecta and the shock has modified also the shape of the central cold condensation.
This is visible from the drop-shaped, cold density enhancement visible at the centre of the {\it x-y} plane.
The gas which composes such a structure and which lies in its short tail is directed towards the centre of the plane. 
Downstream of the cluster, the velocity field shows a hint of alignment perpendicular to the symmetry axis. 
On the right side of the plots and downstream of the shock, in a small part of the plane still unaffected by the shock,
the velocity field still preserves its original features and points towards the cluster centre.

At later times ($t\sim 39Myr$), the shock has entirely crossed the cluster and
a narrow, dense and cold tail has emerged. 
Such a tail started to grow as the gas ram-pressure-stripped from the head of the bow shock was channelled
behind the cluster centre. 
The tail is stretched downstream, and it appears as a noticeable overdensity extending along the direction of motion.
This overdensity is the typical by-product of a massive perturber moving through a uniform medium and creating a density wake
(e. g. Park \& Bogdanovic 2017 and references therein for a recent study on the role of this density wake in case
of dynamical friction in a gaseous medium). 

The motion of the channeled matter flowing around the core is still outlined by the velocity field. 
The gas lying along the tail is accelerated towards the centre of the plane, 
i. e. the {\it x-}component of the velocity vector appears reverted with respect to the gas lying above and below the tail. 
Globally, the stellar distribution appears to follow the elongated shape  of the gas flow, in that
star particles are distributed and scattered around the cluster but tend to gather downstream along the tail, 
forming a drop-like shape. However, the stellar component  
is still strongly condensed in the very centre, as outlined by the thick, orange contours in the density map, which include the large 
majority of the stellar mass (for further detail, see the caption of Fig.~\ref{fig_maps}). 

At this stage, the tail has grown and evolved into  an 'accretion column'
(Bondi \& Hoyle 1944; Shima et al. 1985; Moeckel \& Troop 2009), through which matter can continuously flow onto the cluster
and mix with the AGB ejecta.

The density appears to vary along the tail and its body is composed of several entangled, thin filaments.
Inside the tail, the velocity field presents significant differences from the regular 
behaviour visible elsewehere in the plane.
Globally, the motion of the cold gas composing to the tail is still directed towards the centre, but now it appears chaotic and turbulent,
in particular by looking at the {\it y-}component of the velocity, which in some cases is positive and in other cases negative.

Upstream of the cluster  the gas distribution presents a wide lobe, located below its centre and
protruding towards the bottom-left corner.   
This lobe results from the gravitational effect of the central mass condensation of the cluster,
which at this time is very dense ($\sim 10^{-16}~g/cm^3$, which correspond to $10^6 M_{\odot}/pc^3$, see 
Fig.~\ref{fig_chem1}), and it can deviate the trajectory of the gas flowing towards it 
from the back of the tail, along the accretion column and towards the core.
In some cases, the flowing gas can move past the centre and orbit around it, thus originating the lobe.

The stellar component in the core appears more diffuse than at $26~Myr$, but the mass outside the innermost $\sim~3~pc$ is negligible. 

Little differences are visible in the maps computed at $63~Myr$, i.e. at the end of the star formation. 
The only remarkable features are the disappearance of the lobes and a more circular distribution of the star particles around the core.

\subsection{Evolution of the stellar component}

The left column of Fig.~\ref{fig_chem1} shows the density profiles of FG (green dash-dotted lines) and SG (black dashed lines) stars computed at various times,
as well as the density
profiles of SG stars with different helium abundances. The stellar density profiles are computed by assuming an origin placed at the centre of the computational box. 
In addition to showing the evolutionary path leading to the multiple populations found at the end of our simulations,
the panels at different times illustrate the properties of multiple populations if the SG star formation episode was truncated at earlier times. 

At $t~\sim~10~Myr$, all the He-rich SG stars which have purely formed out of the AGB ejecta follow 
a very steep density profile. 
At $t=25.7~Myr$ the shock has already crossed the cluster centre, and the central density is slightly larger than the one 
at $t~\sim~10~Myr$. 
The total density is still dominated by He-rich $(Y>0.33)$ stars,
but also stars with a lower He content are present, which 
have formed after the re-accretion shock has passed through the core. 
In particular, one SG stellar component with $0.30< Y \le 0.33)$ is visible and  
characterised by a flatter density profile. 
Also another smaller population of stars with $Y<0.30$ is present, visible at the bottom of the panel. 
At this time the central density reaches values larger than $>10^5~M_{\odot}/pc^3$. 

Later, at $39.5~Myr$, four different populations are present, with different degrees of concentration.
Remarkably, the slope of the density profile shows a dependence on the He content: in general, the lower Y values, 
the flatter the resulting distribution. 
  
At $t=63.5~Myr$ the stars with  $0.27< Y \le 0.30$ dominate the density distribution practically at all radii, 
with the exception of the innermost regions (i. e. at radii $<0.7$ pc) which still sees the predominance of the most He-rich stars. 

At all times, in the innermost 2-3$~pc$ the total stellar density is dominated by SG stars, whereas at larger radii it is dominated by FG stars. 
It is interesting to point out that our simulations predict that the more extreme and less diluted SG population is more spatially
concentrated than the more diluted SG stars. 
A similar trend has been found in the observational study of the spatial distributions of the SG populations in
NGC 2808 by Simioni et al. (2016), who showed that the most extreme SG stars are more concentrated than the intermediate SG and
the FG populations (see also Johnson \& Pilachowski 2012 for a spectroscopic study showing a similar trend in  M13).
The study of the kinematics of the multiple populations of NGC 2808 (Bellini et al. 2015) also suggests more significant differences
between the structural and kinematical properties of the most extreme SG populations and those of the FG and other SG populations. 
In the future, it will be important to study further the long-term dynamical evolution of systems like those emerging from our simulations and explore the possible
dynamical implications of the differences in the chemical abundances of the various SG populations
(see e.g. Dalessandro et al. 2018 and Fare, Webb \& Sills 2018  for the possible differences in the distribution of populations with different helium abundances
in dynamically old clusters).  

In the right column of Fig.~\ref{fig_chem1} we show the distribution of the helium
abundance in SG stars for our low-density model calculated at
different times. These panels nicely illustrate the formation history
behind the chemical properties of the SG stars. The SG population is
initially dominated by the extreme SG forming from undiluted AGB
ejecta. As the SG formation proceeds and pristine gas starts to dilute
the AGB ejecta, all the panels show the development of extended
distributions and the appearence of stars characterised by less
extreme chemical abundances. We point out that discrete population
groups can emerge also in the case of a continous star formation as a
result of the chemical discontinuity introduced by the infall of
pristine gas (see also D'Ercole et al. 2008). It is worth noting that
the positions of the two peaks of SG stars are qualitatively in agreement
with the ones observed in real clusters such as, e. g., the stellar
populations of NGC 2808 (Piotto et al. 2007), in which three distinct
populations have been found with Y=0.248, Y=0.3 and Y =0.37 (see also
Milone et al. 2015, D'Antona et al. 2016 for the identification of two
more populations in NGC 2808 and their possible origin). 

\begin{figure*}
 \begin{center}
\includegraphics[width = 2.5in]{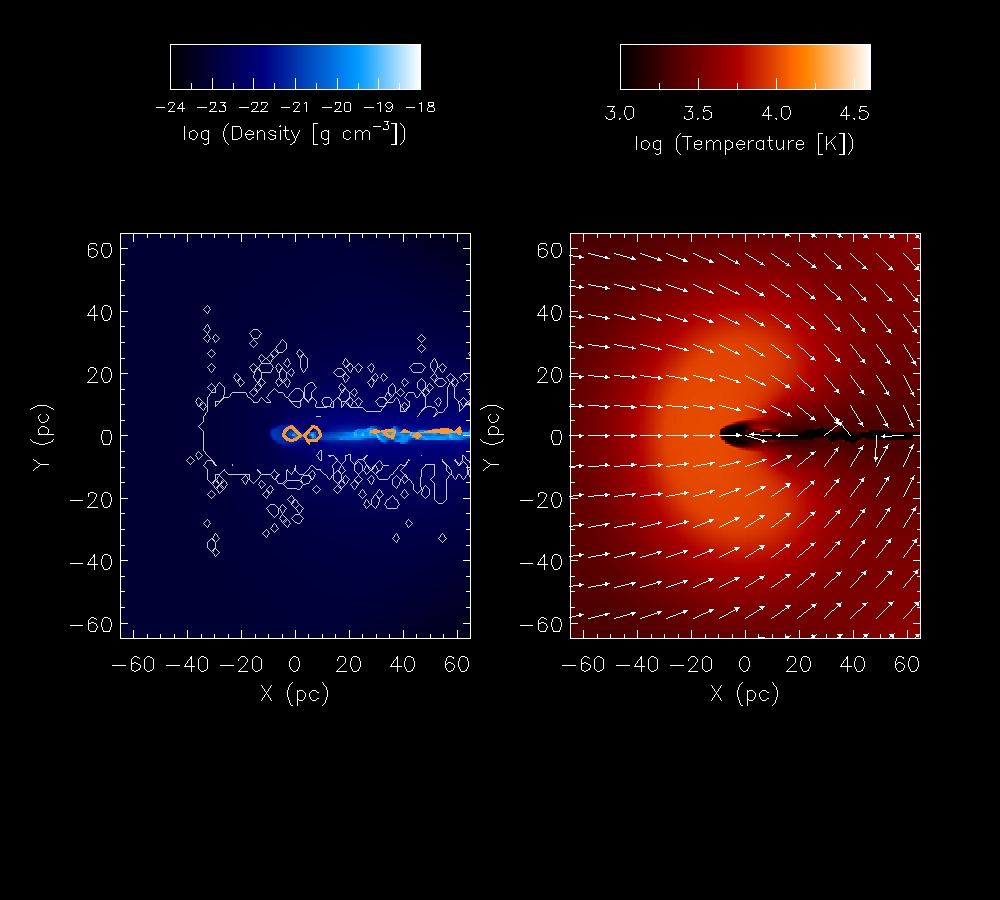}
\includegraphics[width = 2.5in]{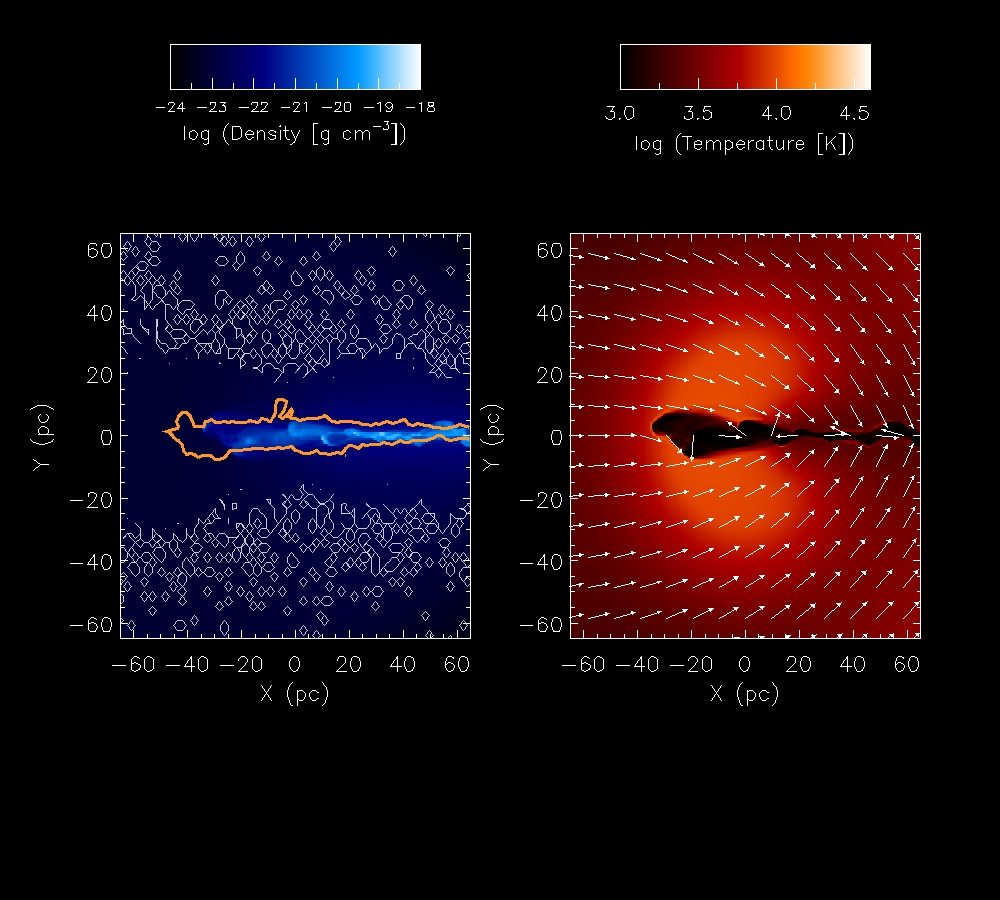}
\includegraphics[width = 2.5in]{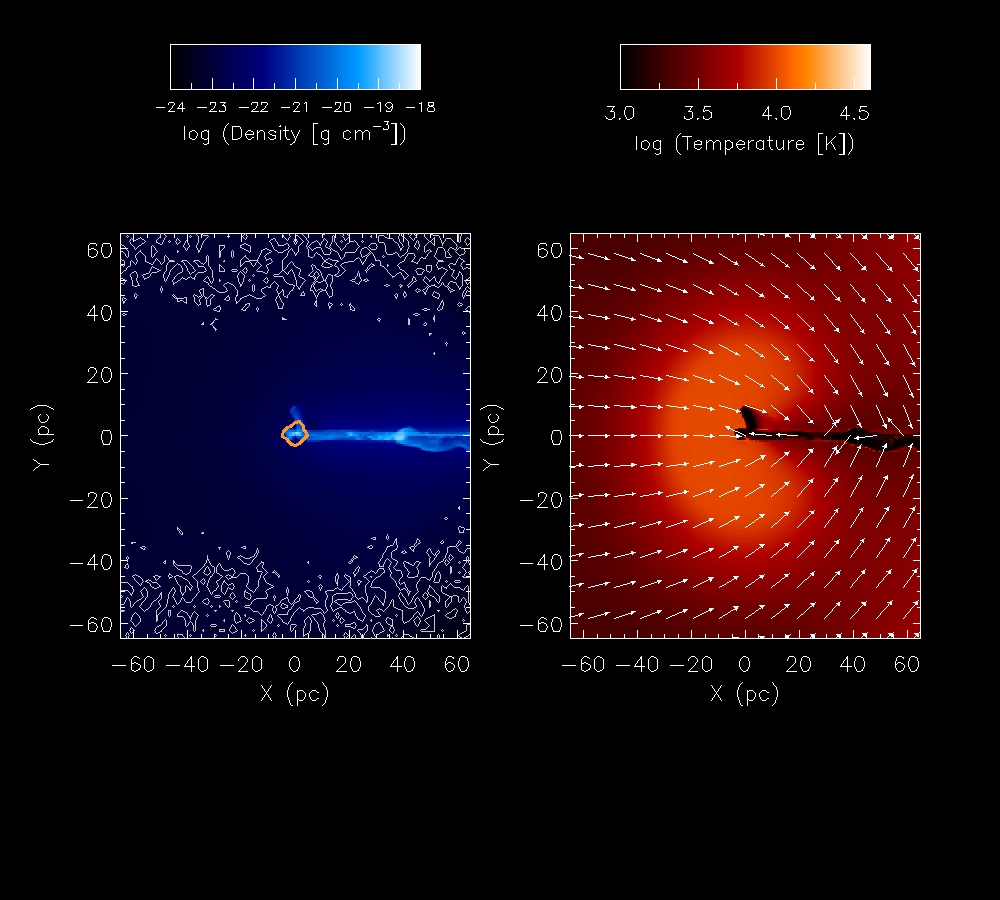}
\includegraphics[width = 2.5in]{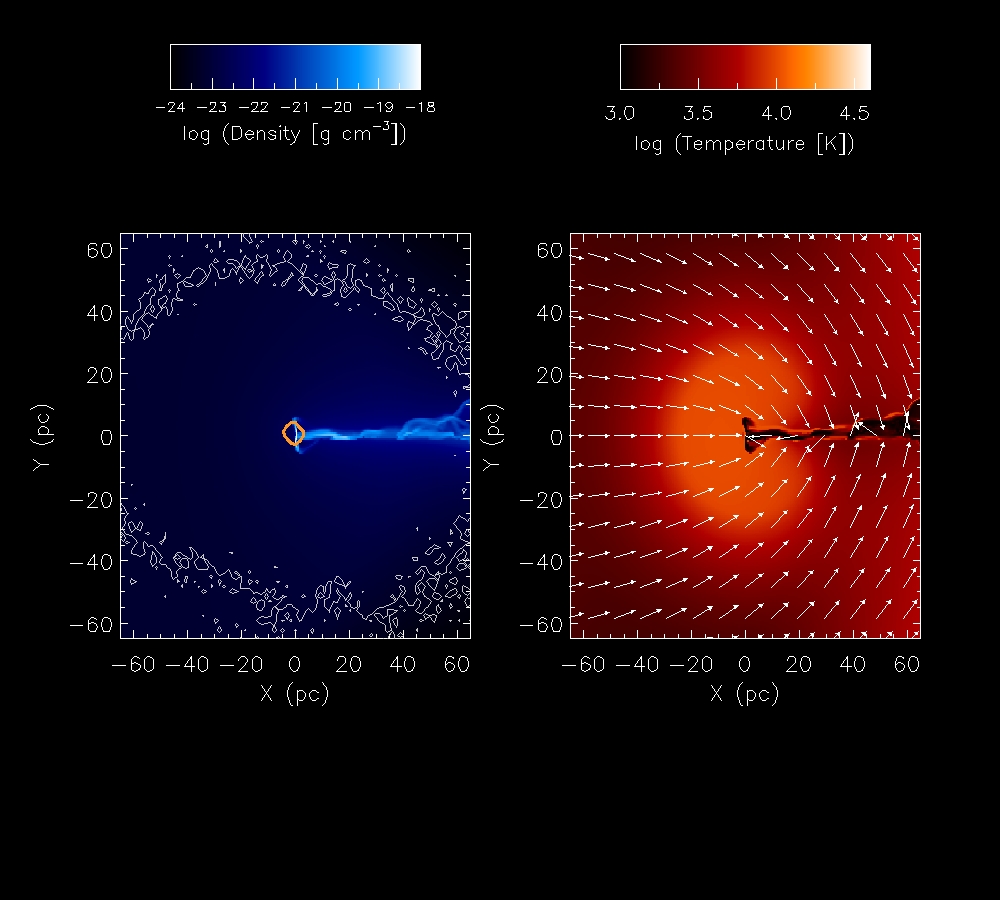}
\end{center}
 \caption{Two-dimensional gas density (left panels of each plot) and temperature (right panels of each plot) maps computed in the {\it x-y} plane for our denser gas simulation at various timesteps
   (from top-left, clockwise: 6, 14, 26 and 64 Myr). 
The white arrows in the temperature maps represent the velocity field.
The white and orange contours describe regions where the SG stellar density is $\ge~10^{-6}$ and $\ge~0.01$ times the maximum value, respectively. The orange contours 
enclose large fractions ($>70 \%$) of the SG stellar mass.} 
\label{fig_maps_densew}
\end{figure*}

\begin{figure*}
   \includegraphics[width=0.7\textwidth]{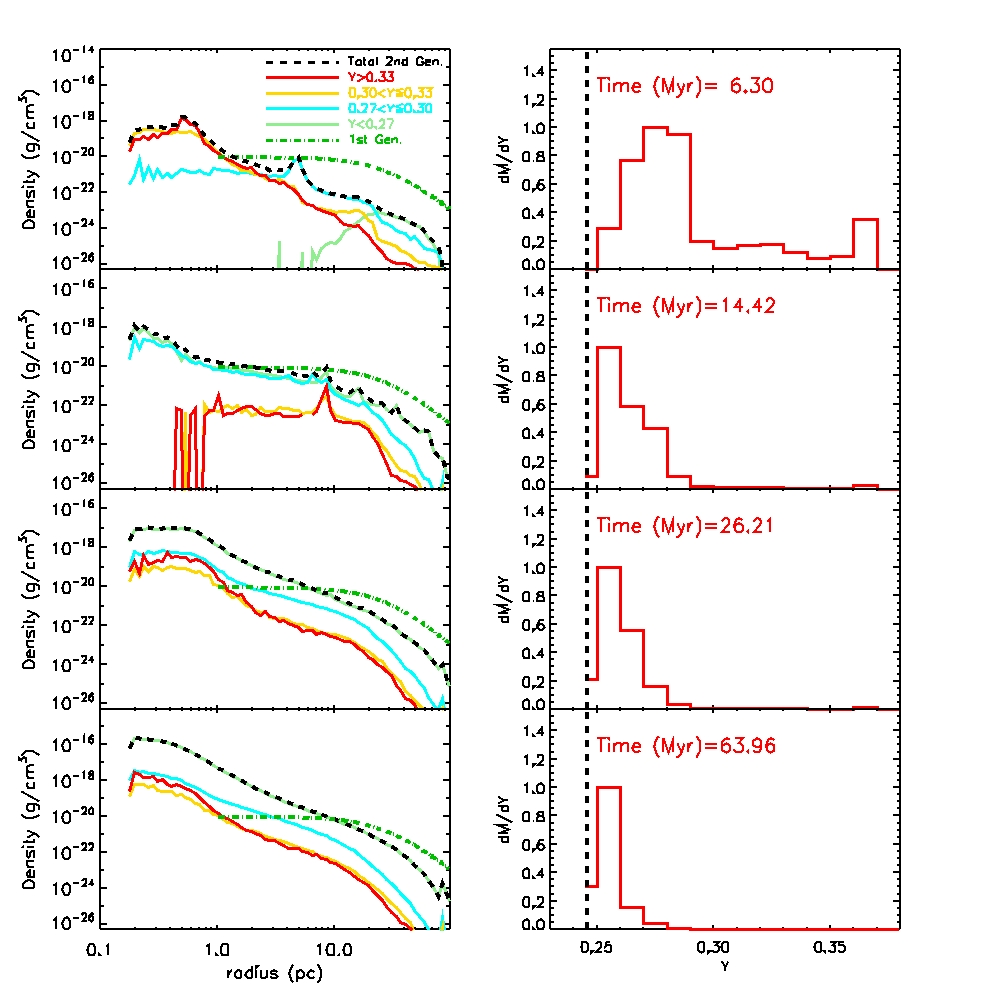}
   \caption{Left column: density profiles of SG stars in our denser model ($\rho=10^{-23}~g/cm^3$) at various timesteps.
     Right column: He mass distribution in SG stars at different times. Lines are as in Fig.~\ref{fig_chem1}.}
        \label{fig_chem1_densew}
\end{figure*}

\subsection{High-density model}

In the case of pristine gas with a density ten times larger ($\rho_{pg}=10^{-23}~g~cm^{-3}$)
than that considered in the previous section, the recollapse of the ISM onto the cluster starts earlier
than in the low-density case, namely 10 Myr 
after the end of the explosions of FG type II SNe, or $\sim$1 Myr after the beginning of the AGB phase. 
This has important consequences on the star formation history of SG stars and on their chemical abundance patterns.

In Fig.~\ref{fig_maps_densew}, we show density and temperature maps of our simulation which includes a denser infalling gas 
in the {\it x-y} plane computed at 6, 14, 26 and 64 Myr, along with contour maps showing the spatial distribution of
SG stars.

It is interesting to point out the significant difference between the structure of the tail and that shown in Fig.~\ref{fig_maps},
in particular the overall inhomogeneous appearance and the presence of various gas condensations visible along the tail. 
The density and temperature maps at 6 Myr already show the presence of a narrow, cold, dense tail along
which various dense knots of gas are recognizable. 
The larger density of the gas is the cause of a stronger cooling occurring along the stream, 
which gives rise to a less homogeneous tail than in the previous case and to the development of more
prominent condensations which are visible when comparing the density maps of the high-density models with those shown 
in Fig.~\ref{fig_maps}.

Overall, the distribution of the stars runs along the stream, as outlined by the thin, white contours.  
In the centre of the density map computed at 6 Myr two very close, nearly circular stellar condensations are visible (thick orange contours in Fig.~\ref{fig_maps_densew}; see its caption for further details),
arising in correspondence of two density enhancements located in the most central region of the plot. 
Other stellar structures are aligned on the symmetry axis of the plot, lying $>20$ pc away from the cluster centre. 
In the temperature map, the motion of the cold gas and its accretion along the tail and onto the cluster is already visible at
6 Myr.

The maps at 14 Myr outline further the various dense gas knots along the stream.
Globally, the extent of the stellar structures appears larger than at the previous time; around the central
region a couple of lobes of cold gas have developed, extrending towards the leftmost boundary and presenting features similar
to those already described when we discussed Fig.~\ref{fig_maps}. 

As vibile in the density map at 26 Myr, the tail appears now more homogeneous than before,
and in the cluster centre most of the new stars seem to have settled onto a more regular, circular distribution. 
The extent and the size
of the lobes seem to have reduced with respect to the previous time.

At 64 Myr the stars are scattered all over in the plane, but 90\% of the total stellar mass
is contained within a very narrow region at the centre of the cluster (orange contour in the density map). 

The presence of various stellar condensations located in different regions of the computational volume 
is confirmed by the stellar density profiles.  In the left column of Fig.~\ref{fig_chem1_densew} 
we show the density profiles of stars of various chemical composition
(as traced by their He content) at different times. 

As in the previous case, the density profiles have been computed in a reference frame with the origin placed
at the centre of the computational grid.

At 6.3 Myr the total SG density profile (black dashed line) presents various peaks associated to some of the 
substructures visible in Fig.~\ref{fig_maps_densew}. 

The distribution of the most He-rich stars presents a peak at $<0.5$ pc, which coincides with the highest
density value of the entire stellar distribution. 
The inner density profile is dominated by the most He-rich (Y$>0.33$) stars, out to radii $\sim 2$ pc.
At radii  $>2$ pc, the highest density values are presented by
stars with  $0.27<$Y$<0.30$, which are clustered between 4 and $\sim$5 pc, 
as also visible from the secondary peak of the total density profile 
at this radius. 
Other two substructures composed of stars with $0.3<$Y$\le~0.33$ and $Y<0.27$ are visible at 20 pc and at radii $>40$ pc, 
respectively. 
The component with the lowest He content dominates the total SG profile at radii $>30$ pc. 

In the SG density profile computed at 14.4 Myr 
the secondary peaks due to the stellar populations born with different compositions
are still partially visible. 

At this time, the two less He-enhanced components contribute similarly to the density profile at radii 
between 0.5 and 10 pc, with a slight predominance of stars with $Y<0.27$, which 
tend to dominate the total SG mass distribution practically at all radii. 

A few distinct, minor peaks are visible in the density profiles of the subpopulations, such as
the one of the most He-rich stars at $\sim 7$ pc.  
The stellar density distributions of the most He-rich components are flatter than the ones seen at 6 Myr. 
  
At $t=26~Myr$ the density profiles appear smoother than at previous times, as they have 
basically lost their original multi-peaked character. As seen earlier, at this time the bulk of the 
stellar component appears regular in shape and occupies the centre of the cluster. 
The bulk of the most He-rich stars occupies the innermost 1 pc.
However, at all radii the SG mass is dominated by the stars with $Y<0.27$.  

These are the main properties of the SG subsystem also at subsequent times,
as confirmed by the density profile computed at the end of the star formation episode ($t=64~Myr$). 

The He mass distributions in SG stars and in the denser model are shown 
in the right column of Fig. ~\ref{fig_chem1_densew}, and they show marked differences with respect to the lower density model.

Also in this case, the panels of Fig.~\ref{fig_chem1_densew} are calculated at various evolutionary times and besides showing the SG formation history,
they also allow us to characterize the properties of the SG if star formations stopped before  $\sim 65$ Myr.

A large separation between the two newly born populations is visible from the 
the Y distributions calculated at 6 Myr. 
The relative heigths of the peaks of
these two populations appear reversed with respect to the case of the low-density model
soon after the beginning of the pristine-gas dilution,  
in that the stars with 
large (Y$\sim$ 0.35) values are subdominant with respect to SG stars with Y$\sim$ 0.27.  

At later times, the Y distribution shows 
a progressive thickening of the less extreme, low-He peak (more similar to the Y value of FG stars) and a
substantial disappearance of the peak of the stars with the highest Y values.

It is interesting to point out that the rapid infall of pristine gas leads to a SG population dominated by stars with
  very modest helium enhancement.

  Finally, an interesting point to discuss concerns the chemical composition of SG stars during the earliest stage of SG formation.
Our results show that at 6 Myr two chemically distinct populations have already formed: a first one rich in He  and a second one characterised
by Y values which are intermediate between the most He-rich SG stars and those of FG stars. 
In terms of mass, it is this second new population which represents the vast majority of the SG stars.
Our results thus indicate that it is possible to have a population with chemical features similar to those of the FG, but formed more than 40 Myr later,
i.e. soon after FG AGB stars start to shed their ejecta. 
The age difference between the latter and the first enriched SG stars made of pure AGB ejecta could be much lower, in this case of the order of 6 Myr. 
Depending on the fraction of FG stars remaining in the cluster, this could complicate the interpretation of observational estimates of age differences
in multiple stellar populations and potentially provide misleading constraints for theoretical models.

\section{Discussion}
\label{sec_disc}
\begin{figure*}
  \includegraphics[width=0.7\textwidth]{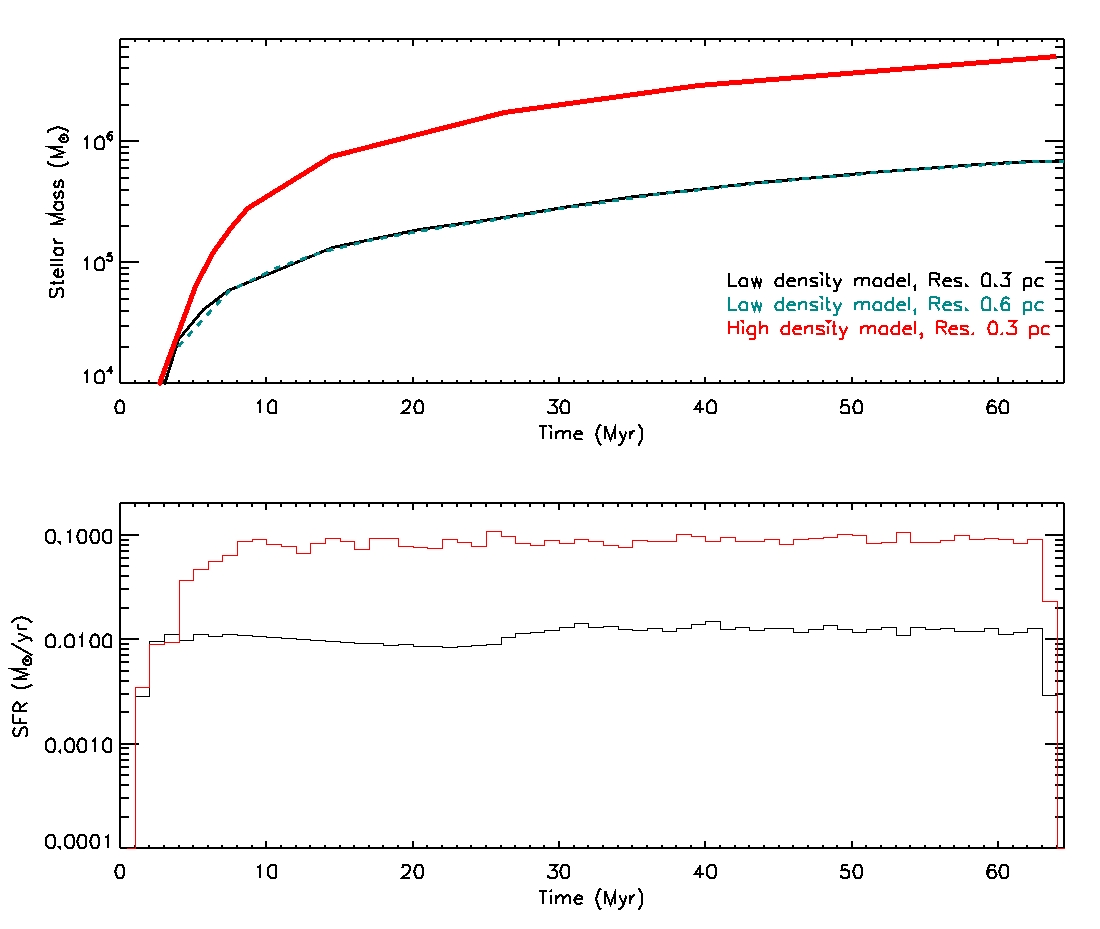}
   \caption{Upper panel: evolution of the SG stellar mass in our simulations.
     The black solid line and the cyan dashed line are the results for the
     low-density model at high and low resolution, respectively.
     The red solid line shows the results in the case of the denser model.
     Lower panel: star formation histories of our low-density and high-density models.
     The black (red) thin solid line is the star formation rate as a function of time for our low- (high-)density
     model. 
   }
        \label{fig_mass}
\end{figure*}

Our simulations have clearly shown how a compact SG subsystem can form in the inner regions of a FG cluster out of the ejecta of AGB
stars along with pristine gas accreted from an external reservoir.
The hydro-dynamical simulations presented here model in detail the dynamics of AGB ejecta and that of the infalling pristine gas; 
our results show that the amount of pristine gas re-accreted and mixed with the AGB ejecta lead to chemical properties in agreement with the main abundance patterns observed in Galactic GCs.
The total stellar mass at the end of our simulations 
amounts to $7\times10^5$ M$_\odot$ for our low-density model, and to  $\sim 5\times~10^6$ M$_\odot$ for the denser model
(upper panel of Fig. \ref{fig_mass}).

Considering that the total initial FG mass in the computational box is $9 \times 10^6~M_\odot$, the ratio of the total initial
FG to the total SG mass formed in the two cases is about 12.9 for the low-density model and 1.8 for the high-density model.  
Considering also that the code does not allow more than 90\% of the gas in each cell to be used for star formation, thus limiting the amount of gas converted in stars as discussed in section~\ref{sec_sf},
these ratios could be slightly smaller, possibly by a factor $\sim~0.9$ (although, owing to the stochasticity of star formation, it is not possible to quantify in detail this effect). 
Although detailed studies of the dynamical evolution of these models are necessary, these results imply 
that the FG mass loss required to produce FG-to-SG mass ratios similar to those observed in old globular clusters 
is significantly less extreme than that often reported in the literature
(suggesting mass loss factors up to 100 or more and referred to as the mass budget problem) and more in line with factors of the order of $\sim$5-10, 
already suggested by previous studies which took into account the contribution of both AGB ejecta and the diluting pristine gas
(see e.g. D'Antona et al. 2013, Ventura et al. 2014). 

In Fig. \ref{fig_mass} we also show the stellar mass as a function of time computed in the low-density model at lower resolution
(0.6 pc). Even if we do not resolve the Jeans scale (as this would require prohibitive computational resources),  
the comparison between the results in the low-density model at different resolution indicates 
that in our simulation numerical convergence is satisfactory.

\subsection{On massive stars in the second generation}
\label{sec_ms}
In our model, an efficient gas accretion and a prolonged star
formation history (shown in the lower panel of Fig.~\ref{fig_mass}) are possible only by neglecting the feedback
of stellar winds and SNII from SG massive stars (i.e. those with $M>8~M_{\odot}$), 
which could in principle hamper mass accretion onto the cluster (Matzner \& Jumper 2015) and,
under certain conditions, even cause the immediate end of star
formation (see D'Ercole et al. 2008 also for the possible role of
other feedback sources).  In addition, should a significant fraction
of SNII ejecta be retained in the cluster they could produce a spread
in, for example, Fe which is found only in a subset of the sample of
Galactic clusters studied so far (see e.g., Renzini et
al. 2015, Da Costa 2016 and references therein).
The possible role of supernovae in the origin of the more complex
manifestations of multiple populations such as the multiple
populations in NGC 2808 has also been discussed by D'Antona et
al. (2016). 

A detailed study of the possible role of massive stars in the dynamics
of the gas during SG star formation and the implications for
the resulting SG chemical properties requires additional simulations
and further study beyond the scope of this investigation. 

Here however we discuss and review a few solutions and interesting
possibilities concerning the role of massive stars in the formation of
the SG population. 

A possibility raised in a few previous studies (see e.g. Decressin et
al. 2007, D'Ercole et al. 2008, 2010) is that the stellar IMF of the SG is
truncated at masses smaller than $M\sim 8-10~M_{\odot}$. As previously pointed out
in the literature (see e.g. Renzini et al. 2015), the process of star 
formation in a pre-existing crowded stellar environment is unexplored 
and it is unclear whether and how the extreme environment of SG 
formation can affect the IMF of the SG.  
Very recently, by means of non-turbulent smoothed particle hydrodynamics simulations, 
Bekki (2019) showed that 
the growth of large, high-density gas clumps that lead to the
formation of massive stars with $M> 8~M_{\odot}$ can be severely
suppressed in GCs. Further studies on both theoretical and observational sides will be needed to
confirm this result and shed more light on the IMF of multiple populations in GCs. 

Another interesting possibility is opened by the extremely compact
structure of the SG subsystem emerging from the simulations presented in this paper.
Such a SG subsystem is characterised by  a 
half-mass radius equal to $0.4~pc$ in the low-density model and of 
$3~pc$  in the high-density model, with a total mass of $\gtrsim$ $ 10^6~M_{\odot}$. 
In such a compact and high-density SG subsystems, if massive
stars form, they are likely to be significantly affected by close
encounters and physical collisions with other SG stars. 
In the case of the more compact SG subsystem emerging from our simulations, 
massive stars could rapidly segregate 
in the inner regions and populate an even denser sub-component before they can 
evolve and explode as SNII.

As for the
high-density model, it is interesting to point out its complex
structure during the very early stages of SG formation, characterised
by a pronounced tail and several small and dense clumps eventually merging to
form the final SG subsystem (see e.g. McMillan, Vesperini \& Portegies
Zwart 2007 for a discussion of the early dynamics and mass segregation
in clumpy systems). 

All the aspects concerning the 
dynamical evolution and the possible collisional history of massive
stars in SG subsystems will require specific simulations which we
defer to a future study; here, however, it is worth mentioning that an
interesting possibility in the context of studies of
multiple-population cluster formation has been proposed by Sills \&
Glebbeek (2010), who suggested that massive star collisions (even
though in their work the massive stars considered belonged to the FG
population) might release processed gas out of which SG stars would
form. In the context of our model, additional gas for the SG formation
might therefore be provided by the physical collisions of the massive
SG stars themselves. In a future study on the early
dynamics of the SG subsystem we will further explore also this aspect.

\subsection{On the detectability of globular clusters during SG star formation}
In our model, the second generation forms at the centre of the first-generation cluster (for a detailed study of the detectability of GCs 
during FG star formation, see Pozzetti et al. 2019) as an extremely compact
and dense stellar subsystem surrounded by a 
more spatially extended first-generation component.
Under these conditions, during the SG formation the cluster should be composed of a faint and nucleated core, possibly surrounded by a diffuse component. 
If the SG contains a few massive stars, they should be visible before possibly undergoing the rapid sequence of collisions facilitated by the cluster dense
environment, 
as discussed in the previous subsection. As new massive stars (MS) form and before suffering collisions with other massive stars, during their drift towards the very centre
they should emit UV light.
To estimate the UV magnitude associated with massive stars in the SG subsystem, one could assume that the most massive stars
with mass $M>M_{dis}$ disappear as they are rapidly affected by collisions in the centre, and that new MS are continuously formed. 

In such a case, we can perform an attempt to estimate the apparent magnitude of the cluster by means of the spectrophotometric code STARBUST99
(Leitherer et al. 2014) and assuming a truncated IMF. 
For this computation, we consider the redshift range probed by the MUSE instrument (Bacon et al. 2017), mounted on the Very Large Telescope, 
and its detected Lyman $\alpha$ emitters ($3\le z \le 6.7$;  e. g., Inami et al. 2017). 
This instrument has allowed considerable steps forward in the identification of extremely faint sources at high $z$,
in particular in gravitationally lensed fields (Karman et al. 2017, Vanzella et al. 2017a,b, Mahler et al. 2018).

When SG stars start to form, we have a $10^7~M_{\odot}$ FG which is already 30 Myr old, which we assume have originated in an instantaneous
burst.

\begin{figure*}
   \includegraphics[width=0.9\textwidth]{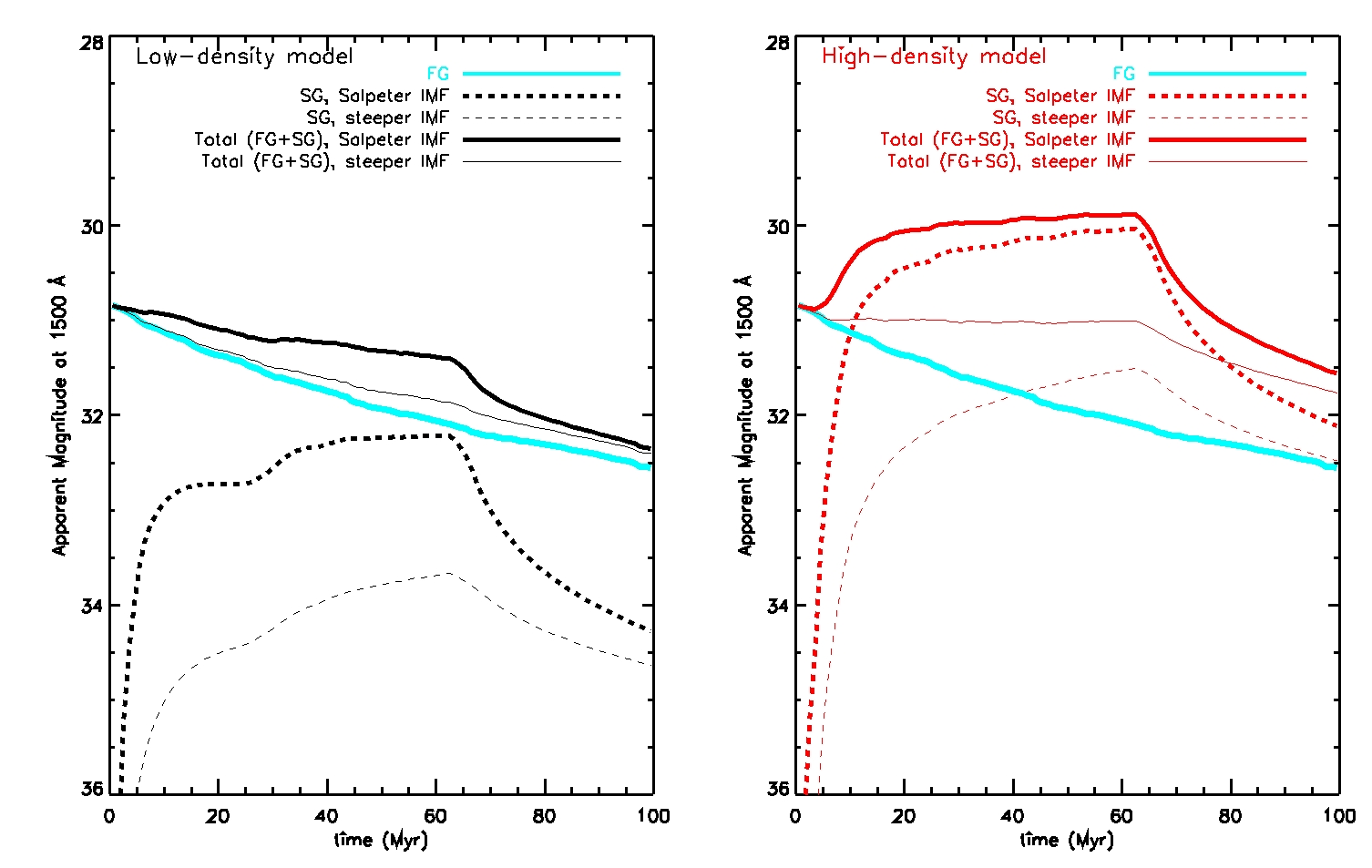}
   \caption{Evolution of the apparent magnitude at 1500 \AA~ of FG and SG stars in the low-density (left panel)
     and high-density (right panel) model, calculated assuming a cluster placed at $z=6$.  
     In each panel, the thick cyan line is the magnitude calculated for FG stars.
     The thick solid and thick dashed black (red) lines are the apparent magnitudes of the entire cluster (including FG and SG stars) and
     of the SG computed with STARBURST99 assuming a Salpeter IMF in the low-density (high-density) model.
     The thin solid and thin dashed black (red) lines are the magnitudes of the entire cluster and of SG stars, respectively, computed assuming a steeper
     IMF, characterised by a slope $\alpha=3.3$, in the low-density (high-density) model.}
        \label{fig_mag}
\end{figure*}

To compute the magnitudes, we have considered the star formation histories shown
in the bottom panel of Fig. \ref{fig_mass} and 
two different IMFs: the Salpeter (1995) IMF, characterised by
a power law with exponent $\alpha=2.35$, and a steeper IMF with $\alpha=3.3$, representative of a 
SG deprived in massive stars.  With these ingredients, we can compute 
the evolution of the 1500 $\AA$ apparent magnitudes of the cluster in all of its components (FG and SG)
in the low-density and high-density models, as shown in Fig.~\ref{fig_mag}. 
In this case, for the computation of the apparent magnitudes the cluster has been placed to $z=6$. 
In the low-density model, SG stars are always fainter than FG stars.
In the case of a Salpeter IMF, due to the effect of SG stars at the peak of their luminosity, occurring at $\sim~60~Myr$,
the cluster is less than one magnitude brighter than the FG component.
On the other hand, in the denser model the luminosity of the cluster is dominated by SG stars
already after a few Myr in the case of a Salpeter IMF, and after 60 Myr in the case of a steeper IMF. 
Even in the case of the low-density model, 
the cluster presents magnitudes not far from those already accessible in current observational samples, in particular 
in lensed fields (Vanzella et al. 2017a). 
Very recently, Vanzella et al. (2019) reported on
a superdense and  compact star-forming region at z=6.143, with an effective radius of 13 pc (with F105W magnitude of 31.1), 
embedded in a more diffuse, larger structure of a size of $\sim~40$ pc (see also Bouwens et al. 2018). 
This object is part of a larger structure, 
which contains several distinct, tiny sources, with magnitudes between 29.6 and $>32$,  
and distributed across a region which extends for several tens kpc. 

These observational studies show that the detection of young GCs are highly facilitated when they are located in lensed fields,
in particular in strongly magnified regions, where very faint magnitudes and small scales can be probed, with good control
of the systematics (Vanzella et al. 2017a,b). 
In such generally narrow regions, the magnification due to strong lensing can reach extremely 
high values ($\mu>10$, see Vanzella et al. 2017a; 2017b) and the magnitudes are boosted by $2.5~log10(\mu)$.

The appearance of the system would be significantly dependent also on its dynamical evolution. 
If a significant fraction of the first generation had expanded, in principle it could be possible to detect the SG as a faint point-like source
surrounded by a diffuse component.

Observationally, the scales of the SG will be addressable in the future thanks to the ESO Extremely Large Telescope which,
if exploited in lensed fields, will allow us to spatially resolve systems of $\sim~2~pc$ at $z>3$ and down to very faint magnitudes, ($>32$, 
Fiorentino et al. 2018).

Other considerations related to the environment of these objects are in order. 
In principle, one could also expect the high redshift precursors of GCs to be clustered at their birth as in the local Universe 
(Kravtsov \& Gnedin 2005; Renzini 2017), i.e. to be aggregated into groups of newly formed systems. 
This seems to be confirmed by a new study of star-forming complexes in a lensed field at $z\sim 6$, 
which has shown the presence of a handful of very faint sources lying within a group of approximate size of less than 10 arcsec, corresponding to $\sim 60$ kpc (Vanzella et al. 2019).

These system are also expected to be surrounded by a large reservoir of cold gas, perhaps visible as a diffuse 
Lyman $\alpha$ component. In fact, due to its resonant nature, the $Ly\alpha$ transition is an extremely powerful 
tracer of the structure of the neutral gas located around stellar structures (e. g. Gronke et al. 2017).

A large reservoirs of dense, cold gas is 
required to fuel further SF episodes, and it has to be readily available nearby the systems (D'Ercole et al. 2016). 
If present in large amount and dense enough, this reservoir of cold gas could also be detectable by means of strong infrared lines, such as
the [CII]158$\mu m$ fine structure line,  
a very efficient and dominating coolant for neutral gas.

\subsection{Future directions}
Beside investigating further the structural parameters of the cluster and the ones related to the properties of the external environment, 
in the future it will be important to extend the present study to explore 
other scenarios proposed for multiple populations in GCs. 
  
So far, the origin of multiple populations has generally not been addressed in 
hydrodynamical simulations of globular cluster formation (e. g., Nakasato et al. 2000; Fujii \& Portegies Zwart 2016).
One exception is the study of Howard et al. (2019), which use 3D radiative hydrodynamics simulations to study the formation of 
of young massive clusters and their subsequent evolution for 5 Myr. 

It will be important to exploit such models to study the star formation and the dynamical
evolution of young clusters for more extended evolutionary times.
Currently, this is computationally challenging for a number of reasons. 
One difficulty is represented by the necessity to include massive stellar feedback, 
in both pre-SN and SN phases. Previous studies have shown that the energetic feedback
of MS can accelerate their ejecta and the surrounding gas to very high velocities ($>1000~km/s$),
which can lead to extremely short computational timesteps, which are necessary to 
satisfy the Courant-Friedrichs-Lewy condition (Emerick et al. 2019; Romano et al. 2019). 
Moreover, the  dynamical evolution of massive stars need to be treated by means of direct N-body simulations, which are
generally problematic for large number of particles ($>500000$ stars, Heggie 2014). 
Steps forwards are currently being attempted with new parallel computing technologies (e. g. GPUs) for N-body studies,
but the problem of the coupling with hydrodynamics is still to be tackled.\\
To test scenarios such as the one of massive (Elmegreen 2017) and supermassive stars (Gieles et al. 2018), 
more physical ingredients will need to be included in hydrodynamical simulations, including fully  collisional  N-body  dynamics. 
A new numerical method to resolve the dynamics of the stars and gas which include
coupling of hydrodynamic codes, an N-body code and a stellar evolution code is the 
Astro-physical MUlti-purpose  Software  Environment  (AMUSE, Pelupessy et al. 2013; Portegies Zwart \& McMillan 2018).
Recently, Wall et al. (2019) presented an interface for the FLASH hydro-code in the AMUSE framework.
Such a work will be a reference for any future attempt to model multiple populations in young stellar clusters with 
hydrodynamic codes in various scenarios. 

\section{Conclusions}
In this paper we have used grid-based, three-dimensional hydrodynamical simulations to model the formation of second generation stars in a massive cluster moving
through a uniform background gas distribution. After the explosion of all FG type II SNe,  when the cluster is about 40 Myr old, 
second generation stars start to form out of the  ejecta of AGB stars 
and external gas with pristine (i.e. same as FG stars) composition accreted from the environment in which it is moving. The epoch of formation of second-generation stars is assumed to last
about 60 Myr, until the cluster is $\sim$100 Myr old. 
Our study is focused on a massive cluster with an initial first-generation mass equal to $10^7 M_{\odot}$, 
($9 \times 10^6 M_{\odot}$ inside the computational box considered in our hydro simulations) 
a possible progenitor of the most massive clusters in the old Galactic GC population. 
We do not model the formation of first generation stars and assume a cluster already in place,
and that at an age of 40 Myr all the primordial gas and the FG type II SNe ejecta have been expelled. 
The occurrence of type II SNe has left the cluster in an extended cavity, which is then replenished by the later recollapse of the interstellar gas onto the cluster.
We have considered two different values for the density of the background pristine gas, a low-density case with $\rho=10^{-24} g/cm^3$ and a high-density one with $\rho=10^{-23} g/cm^3$. 
These two choices lead to different infall times, with important implications for the resulting chemical abundance pattern of the SG population. 
Although some of the conditions chosen for our model are idealised, the physical parameters chosen for the inflowing matter 
have been selected in order to describe gas densities observed in real systems and expected 
to be typical for star-forming galaxies at high redshift. Our main results can be summarised as follows. 
\begin{itemize} 
\item In the low-density model, a very compact, dense SG stellar subsystem starts to form at the cluster centre immediately at the beginning of our simulation
from the undiluted ejecta of the most massive AGB stars. The low density of the external gas causes the re-collapse of the gas in the cavity,
and the resulting dilution, to impact on the centre of the cluster approximately 20-25 Myr after the onset of star formation.
After this time, the interstellar gas flows towards the centre of the system mostly through an accretion column, which has emerged soon after a turbulent
tail aligned with the axis of motion was formed. 
The SG formed in this simulation is characterised by two distinct groups: one with strong helium enhancement formed from undiluted AGB ejecta and one with more moderate helium enhancement formed later in the SG star formation event from AGB
ejecta diluted with pristine gas. 
\item In the denser gas model, the pristine gas infall is more rapid and therefore the dilution of the AGB ejecta forming SG stars starts earlier.
This implies that the SG population is dominated by stars with less extreme chemical anomalies, as traced by the He abundances of SG stars and as observed in many Galactic GCs.
In this case the tail along which gas can be accreted presents a more clumpy structure, and in correspondence of the clumps new separate stellar aggregates can form,
located in different positions in the computational volume. Approximately 26 Myr after the start of the SG star formation the morphology of the cluster appears
more regular and approximately spherical. Also in this case, the SG subsystem is concentrated in the innermost regions of the cluster.
\item  We find that the first and most helium-enhanced SG stars are more centrally concentrated than SG stars
 born later and with less extreme He content. This trend is consistent with a few studies
 in which the spatial distribution of the different SG subpopulations was investigated (see e.g. Simioni et al. 2016, Johnson \& Pilachowski 2012). 
\item
  In the two models presented here, at the end of the SG formation the ratio of the initial mass in FG stars to that in SG stars is equal to about 1.8
  for the high-density model and about 12.8 for the low-density model.
  Although these values are larger than the observed present-day values of this ratio (see e.g. Milone et al. 2017)
  and require a subsequent preferential loss of FG stars, such a required FG mass loss is significantly less extreme than that often reported in the literature.
Previous studies have explored the possible dynamical paths leading to the needed loss of FG stars during the cluster early and long-term dynamical evolution (see e.g. D'Ercole et al. 2008) and we will further extend the study of this dynamical aspect in future investigations.
\item In both models presented here, already after 10-20 Myr the SG subsystem is very compact and characterised by a high cental density ($>10^5 ~M_{\odot}/pc^3$).
Although further studies to explore in detail the early and long-term dynamics of the SG subsystem are necessary, we pointed out that the
high density of the SG subsystem found in our simulations may lead to rapid and numerous physical collisions between massive SG stars before they explode as SNII.
An interesting consequence of these collisions might be the stripping of the stars' outer layers possibly providing additional gas for the formation of SG stars.
\item By means of the STARBURST99 code, we have computed the expected apparent magnitude of the simulated cluster, assuming that it is placed at $z=6$,
as some recently-discovered high-redshift young stellar clusters, accessible thanks to the MUSE instrument.
We have tested two different choices for the stellar initial mass function, 
including a standard, Salpeter (1955) power-law IMF and a steeper IMF, representative of a SG deprived in massive stars. 

Even in the most conservative, low-density model, the cluster presents UV magnitudes between $\sim $31 and 32,
not far from those already observationally accessible 
in lensed fields (Vanzella et al. 2019).  
\end{itemize} 
In the future, we plan to extend the study presented here with a survey of simulations aimed at exploring the SG formation in clusters with a broader range 
of different structural parameters and at following their subsequent early and long-term dynamical evolution.
The analysis presented here is conducted in the AGB scenario proposed by D'Ercole et al. (2008).
Of course, all the other proposed models might lead to a different evolution of the chemical and structural
properties of GCs. The extension of the simulations presented here to these scenarios will be therefore valuable
to allow a better discrimination between various hypotheses proposed so far for the formation of GCs. 

\section*{Acknowledgments}
An anonymous referee is acknowledged for several useful comments. 
We are grateful to Emanuele Dalessandro, Franca D'Antona, Eugenio Carretta and Sandro Bressan for interesting discussions.
This research was supported in part by Lilly Endowment, Inc., through its support for the Indiana University Pervasive Technology Institute, and in part by the Indiana METACyt Initiative.
The Indiana METACyt Initiative at IU is also supported in part by Lilly Endowment, Inc.
F.C. acknowledges funding 
from the INAF PRIN-SKA 2017 program 1.05.01.88.04.



\end{document}